\documentclass[onecolumn,nofootinbib]{revtex4}
\usepackage{setspace}
\usepackage{amsmath,amssymb}
\usepackage{graphicx}
\usepackage{dcolumn}
\usepackage{bm,color}
\usepackage{hyperref}
\usepackage{accents}
\usepackage{amssymb,float}
\usepackage{amsmath}
\usepackage{multirow}
\usepackage{tikz}
\usepackage{enumerate}

\hypersetup{
    colorlinks=true,
    linkcolor=red,
    filecolor=magenta,      
    citecolor=blue
}

\newcommand{\udt}[3]{#1^{#2}_{\phantom{#2}#3}}
\newcommand{\udut}[4]{#1^{#2\phantom{#3}#4}_{\phantom{#2}#3\phantom{#4}}}

\newcommand{\dut}[3]{#1_{#2}^{\phantom{#2}#3}}
\newcommand{\dudt}[4]{#1_{#2\phantom{#3}#4}^{\phantom{#2}#3}}
\newcommand{\lc}[1]{\accentset{\circ}{#1}}%Levi-Civita connection

\begin{document}

\title{Classification of Teleparallel Horndeski Cosmology via Noether Symmetries}

\author{Konstantinos F. Dialektopoulos}
\email{kdialekt@gmail.com}
\affiliation{Laboratory of Physics, Faculty of Engineering, Aristotle University of Thessaloniki, 54124 Thessaloniki, Greece}
\affiliation{Center for Gravitation and Cosmology, College of Physical Science and Technology, Yangzhou University, Yangzhou 225009, China}

\author{Jackson Levi Said}
\email{jackson.said@um.edu.mt}
\affiliation{Institute of Space Sciences and Astronomy, University of Malta, Malta, MSD 2080}
\affiliation{Department of Physics, University of Malta, Malta}

\author{Zinovia Oikonomopoulou}
\email{zhnobia.oikonomopoulou.21@um.edu.mt}
\affiliation{Institute of Space Sciences and Astronomy, University of Malta, Malta, MSD 2080}

\date{\today}

\begin{abstract}
Teleparallel Horndeski theory offers an avenue through which to circumvent the speed constraint of gravitational waves in an efficient manner. However, this provides an even larger plethora of models due to the increase in action terms. In this work we explore these models in the context of cosmological systems. Using Noether point symmetries, we classify the dynamical systems that emerge from Teleparallel Horndeski cosmologies. This approach is very effective at selecting specific models in the general class of second-order Teleparallel scalar-tensor theories, as well as for deriving exact solutions within a cosmological context. By iterating through the Lagrangians selected through the Noether symmetries, we solve for a number of cosmological systems which provides new cosmological systems to be studied.
\end{abstract}

\maketitle

\section{Introduction}\label{sec:intro}

General relativity (GR) as the fundamental theory that describes gravity in the standard model of cosmology ($\Lambda$CDM) has seen overwhelming successes in describing the evolutionary processes in the Universe \cite{misner1973gravitation,Clifton:2011jh,Aghanim:2018eyx}. In this scenario, the current phase of expansion is being driven by the vacuum energy associated with spacetime \cite{Riess:1998cb,Perlmutter:1998np}. However, fundamental issues remain prevalent for a cosmological constant $\Lambda$ scenario \cite{RevModPhys.61.1,Appleby:2018yci,Ishak:2018his}. Along a similar vein, the prospect of direct observations of cold dark matter (CDM) particles remains elusive \cite{Baudis:2016qwx,Bertone:2004pz}. More recently, $\Lambda$CDM has been met by challenges from observational cosmology in the form of the Hubble tension \cite{Bernal:2016gxb,DiValentino:2020zio,DiValentino:2021izs,Riess:2021jrx} where a disagreement in the value of the Hubble constant has appeared between late-time cosmology-independent measurements of $H_0$ \cite{Riess:2019cxk,Wong:2019kwg} in contrast to observations from the early Universe which are then used to predict the value of $H_0$ using $\Lambda$CDM \cite{DES:2017txv,Aghanim:2018eyx}.

In order to better meet these challenges, a possible solution may be to reconsider GR as the fundamental description of gravitation within the standard model of cosmology \cite{Clifton:2011jh,Capozziello:2011et,CANTATA:2021ktz,Nojiri:2010wj}. The possible directions that theories modified beyond GR can take has branched out into several directions over the years with the Lovelock theorem as one key guide \cite{Lovelock:1971yv} in these studies. One of these proposals is the addition of a single nonmimally coupled scalar field to the Einstein-Hilbert action which has led to the umbrella of Horndeski gravity \cite{Horndeski:1974wa} which is dynamically equivalent to most modified theories of gravity in curvature-based scenarios. Horndeski gravity is a rich arena for constructing cosmological models \cite{Kobayashi:2019hrl}. However, recent multimessenger signals have put tight constraints on the speed of gravitational waves of at most one part in $10^{15}$ in comparison with the speed of light \cite{TheLIGOScientific:2017qsa,Goldstein:2017mmi}. This has led to a severely restricted form of regular Horndeski gravity that continues to be observationally viable  \cite{Ezquiaga:2018btd}.

This adds a strong motivation to consider possible alternatives that may revitalize the search for a viable model in terms of phenomenological within this framework. One such possibility is the transformation from curvature- to torsion-based theories of gravity \cite{Bahamonde:2021gfp,Aldrovandi:2013wha,Cai:2015emx,Krssak:2018ywd}. In this setting, we consider Teleparallel gravity (TG) which embodies the exchange of the Levi-Civita to the Teleparallel connection \cite{Weitzenbock1923,Bahamonde:2021gfp}, which is torsion-full but continues to satisfy metricity. This represents a transformation in the geometry of the theory and thus the regular measures of curvature identically vanish, such as the Ricci scalar $\lc{R}$ (over-circles represent quantities calculated with the Levi-Civita connection), which in TG gives $R \equiv 0$. By relating both geometries, it can is found that the regular Ricci scalar is dynamically equal to a torsion scalar $T$, up to a boundary term $B$. This guarantees that GR is dynamically equivalent to a \textit{Teleparallel equivalent of general relativity} (TEGR). Thus, in TG the division between the second- and fourth-order contributions to the Einstein-Hilbert action are decoupled which produces a softened version of the Lovelock theorem in this setting \cite{Lovelock:1971yv,Gonzalez:2015sha,Bahamonde:2019shr}.

TEGR can be generalized in a number of interesting ways, using the same reasoning as $f(\lc{R})$ \cite{Sotiriou:2008rp,Faraoni:2008mf,Capozziello:2011et,Nojiri:2017ncd} we can consider $f(T)$ gravity \cite{Ferraro:2006jd,Ferraro:2008ey,Bengochea:2008gz,Linder:2010py,Chen:2010va,Bahamonde:2019zea}, or even $f(T,B)$ gravity \cite{Escamilla-Rivera:2019ulu,Bahamonde:2015zma,Capozziello:2018qcp,Bahamonde:2016grb,Paliathanasis:2017flf,Farrugia:2018gyz,Bahamonde:2016cul,Wright:2016ayu} which have both shown interesting results in the literature. Another interesting generalization of TEGR that has gained interest in the literature is that of using the Teleparallel analogue of the Gauss-Bonnet scalar $T_G$ which can be used to produce $f(T,T_G)$ gravity \cite{Kofinas:2014owa,Bahamonde:2016kba,delaCruz-Dombriz:2017lvj,delaCruz-Dombriz:2018nvt}. A novel approach in this direction has been recently suggested in Ref.~\cite{Bahamonde:2019shr} where a TG analogue of Horndeski theory was developed. Due to the organically lower order nature of TG, this produces a much richer landscape in which to produce scalar-tensor models. This inadvertently also helps revive previously disqualified models since the gravitational wave propagation equation becomes much more intricate \cite{Bahamonde:2019ipm}. Most models in the Teleparallel Horndeski theory also satisfy the parameterized post-Newtonian conditions \cite{Bahamonde:2020cfv} and produce a varied polarization structure for gravitational waves \cite{Bahamonde:2021dqn}. More recently, Teleparallel Horndeski gravity has been studied for its self-tuning properties in Refs.~\cite{Bernardo:2021bsg,Bernardo:2021izq}.

In this work, we aim to use the \textit{Noether symmetry approach} \cite{Capozziello:1996bi} to classify the ensuing models that can be produced within the Teleparallel Horndeski plethora of models. This approach is a vital tool to probing physical models of a landscape of scalar-tensor theories such as Horndeski gravity since it can use symmetries to reduce the complexity of a system of equations \cite{Basilakos:2011rx}. In practice, the method depends on a point-like Lagrangian and a symmetry that leaves the Lagrangian invariant, which are then used to reduce the complexity for a systems so that analytic solutions may potentially be found \cite{Dialektopoulos:2019mtr}. These symmetries are always connected with conserved quantities in a system under investigation. Now, this investigative technique has been used in several settings \cite{Dialektopoulos:2018qoe} such as $f(\lc{R})$ \cite{Capozziello:2008ch,Paliathanasis:2011jq}, scalar-tensor theories \cite{Dimakis:2017zdu,Dimakis:2017kwx,Giacomini:2017yuk,Paliathanasis:2014rja}, as well as non-local theories \cite{Bahamonde:2017sdo}. In TG, this has been used in $f(T)$ gravity \cite{Basilakos:2013rua}, $f(T,B)$ gravity \cite{Bahamonde:2016grb} and $f(T,T_G)$ gravity \cite{Capozziello:2016eaz}. As an approach of classification of theories in regular Horndeski gravity, this approach was presented in Ref.~\cite{Capozziello:2018gms} where only the invariance of the field equations under a Noether point symmetry was considered. This work also led to a number of interesting new solutions. In the present work, we generalize this approach to Teleparallel Horndeski gravity to broaden this classification approach and to obtain novel solutions.

Our study first opens with a brief discussion of TG and the construction of the Teleparallel analogue of Horndeski gravity which takes place in Sec.~\ref{sec:BDLS}. We then discuss our use of the Noether symmetry approach in Sec.~\ref{sec:Noeth_symm} where our point-like Lagrangian is written down together with the Noether symmetries that are then investigated. In Sec.~\ref{sec:models} we use the Noether symmetry approach for the point-like Lagrangian to classify the ensuing classes of models that emerge. Here, we discuss how the regular Horndeski models are complemented by the new additions from this larger form of Horndeski gravity. Finally, we give a summary of our main results in Sec.~\ref{sec:conc}.

\section{Teleparallel Horndeski Cosmology}\label{sec:BDLS}

The curvature associated with GR is recast in TG with the torsional geometric framework offered by the Teleparallel connection which replaces the Levi-Civita connection for gravitational interactions \cite{Aldrovandi:2013wha,Bahamonde:2021gfp,Cai:2015emx,Krssak:2018ywd}. By reconsidering the foundations of GR, TG can offer an alternative approach by which to construct gravitational theories.

In TG the metric tensor is replaced as the fundamental dynamical variable by the tetrad $\udt{e}{A}{\mu}$ and spin connection $\udt{\omega}{B}{C\nu}$, where Greek indices refer to coordinates on the general manifold while Latin ones represent local Minkowski space $\eta_{AB}$. In this way, the tetrads can be used to raise Minkowski indices to the general manifold through \cite{RevModPhys.48.393}
\begin{align}
    g_{\mu\nu} = \udt{e}{A}{\mu}\udt{e}{B}{\nu} \eta_{AB}\,,& &\text{and}& &\eta_{AB} = \dut{E}{A}{\mu}\dut{E}{B}{\nu} g_{\mu\nu}\,,\label{eq:metr_trans}
\end{align}
which also adhere to orthogonality conditions
\begin{align}
    \udt{e}{A}{\mu}\dut{E}{B}{\mu} = \delta_B^A\,,& &\text{and}& &\udt{e}{A}{\mu}\dut{E}{A}{\nu} = \delta_{\mu}^{\nu}\,,
\end{align}
where $\dut{E}{A}{\mu}$ represents the inverse tetrad. Given a metric, there exists an infinite number of the tetrad components that satisfy these relations due to the six local Lorentz degrees of freedom $\udt{\Lambda}{A}{B}$. These degrees of freedom are represented by the spin connection. Together, the tetrad-spin connection pair represented the fundamental variables of the theory.

The Levi-Civita connection $\udt{\lc{\Gamma}}{\sigma}{\mu\nu}$ (to recall, over-circles refer to any quantities based on the Levi-Civita connection) associated with curvature-based theories is replaced in TG with the teleparallel connection $\udt{\Gamma}{\sigma}{\mu\nu}$. This can also be done in GR but is much less common \cite{Hohmann:2021fpr}. The teleparallel connection is curvature-less and satisfies metricity \cite{Weitzenbock1923}, and can be expressed as \cite{Cai:2015emx,Krssak:2018ywd}
\begin{equation}
    \udt{\Gamma}{\lambda}{\nu\mu}=\dut{E}{A}{\lambda}\partial_{\mu}\udt{e}{A}{\nu}+\dut{E}{A}{\lambda}\udt{\omega}{A}{B\mu}\udt{e}{B}{\nu}\,,
\end{equation}
where the spin connection must satisfy \cite{Bahamonde:2021gfp}
\begin{equation}
    \partial_{(\mu}\udt{\omega}{A}{|B|\nu)} + \udt{\omega}{A}{C(\mu}\udt{\omega}{C}{|B|\nu)} \equiv 0\,,
\end{equation}
in order to be the flat connection associated with TG. In this way, the tetrad-spin connection pair balance each other in terms of the freedom of theory. It is important to point out that tetrad frames exist which are compatible with zero spin connection components (as long as they satisfy the corresponding field equations). This is called Weitzenb\"{o}ck gauge \cite{Krssak:2018ywd}.

Teleparallel geometry is based on the replacement of the Levi-Civita with the teleparallel connection, which means that the Riemann tensor identically vanishes ($\udt{R}{\alpha}{\beta\gamma\epsilon}(\udt{\Gamma}{\sigma}{\mu\nu}) \equiv 0$)\footnote{Naturally, this does not mean that the regular Riemann tensor vanishes in general, i.e. $\udt{\lc{R}}{\alpha}{\beta\gamma\epsilon}(\udt{\lc{\Gamma}}{\sigma}{\mu\nu}) \neq 0$.}. Thus, we consider the torsion tensor which is defined through the antisymmetry operator on the connection \cite{Aldrovandi:2013wha,ortin2004gravity}
\begin{equation}
    \udt{T}{A}{\mu\nu} := 2\udt{\Gamma}{A}{(\nu\mu)}\,,
\end{equation}
which acts as a measure of the field strength of gravitation in TG \cite{Bahamonde:2021gfp}, and where the square brackets denote antisymmetric operator. The torsion tensor transforms covariantly under local Lorentz transformations and diffeomorphisms \cite{Krssak:2015oua}, and can be decomposed into irreducible pieces, namely axial, vector and purely tensorial parts given by \cite{PhysRevD.19.3524,Bahamonde:2017wwk}
\begin{align}
    a_{\mu} & :=\frac{1}{6}\epsilon_{\mu\nu\lambda\rho}T^{\nu\lambda\rho}\,,\\
    v_{\mu} & :=\udt{T}{\lambda}{\lambda\mu}\,,\\
    t_{\lambda\mu\nu} & :=\frac{1}{2}\left(T_{\lambda\mu\nu}+T_{\mu\lambda\nu}\right)+\frac{1}{6}\left(g_{\nu\lambda}v_{\mu}+g_{\nu\mu}v_{\lambda}\right)-\frac{1}{3}g_{\lambda\mu}v_{\nu}\,,
\end{align}
where $\epsilon_{\mu\nu\lambda\rho}$ is the totally antisymmetric Levi-Civita tensor in four dimensions. Naturally, when contracted with each other these irreducible parts vanish. This decomposition can also be used to form parts of the torsion scalar invariant through the scalars
\begin{align}
    T_{\text{ax}} & := a_{\mu}a^{\mu} = -\frac{1}{18}\left(T_{\lambda\mu\nu}T^{\lambda\mu\nu}-2T_{\lambda\mu\nu}T^{\mu\lambda\nu}\right)\,,\\
    T_{\text{vec}} & :=v_{\mu}v^{\mu}=\udt{T}{\lambda}{\lambda\mu}\dut{T}{\rho}{\rho\mu}\,,\\
    T{_{\text{ten}}} & :=t_{\lambda\mu\nu}t^{\lambda\mu\nu}=\frac{1}{2}\left(T_{\lambda\mu\nu}T^{\lambda\mu\nu}+T_{\lambda\mu\nu}T^{\mu\lambda\nu}\right)-\frac{1}{2}\udt{T}{\lambda}{\lambda\mu}\dut{T}{\rho}{\rho\mu}\,,
\end{align}
which are parity preserving \cite{Bahamonde:2015zma}. These three scalars form the most general gravitational Lagrangian formed only by parity preserving scalars, $f(T_{\text{ax}},T_{\text{vec}},T{_\text{ten}})$, that is quadratic in contractions of the torsion tensor. Together, the torsion scalar can be formed by
\begin{equation}
    T:=\frac{3}{2}T_{\text{ax}}+\frac{2}{3}T_{\text{ten}}-\frac{2}{3}T{_{\text{vec}}}=\frac{1}{2}\left(\dut{E}{A}{\lambda}g^{\rho\mu}\dut{E}{B}{\nu}+2\dut{E}{B}{\rho}g^{\lambda\mu}\dut{E}{A}{\nu}+\frac{1}{2}\eta_{AB}g^{\mu\rho}g^{\nu\lambda}\right)\udt{T}{A}{\mu\nu}\udt{T}{B}{\rho\lambda}\,,
\end{equation}
which can be shown to be equivalent to the Ricci scalar $\lc{R}$ up to a total divergence term \cite{Bahamonde:2015zma}
\begin{equation}
    R=\lc{R}+T-\frac{2}{e}\partial_{\mu}\left(e\udut{T}{\lambda}{\lambda}{\mu}\right)=0\,,
\end{equation}
where $R$ is the teleparallel connection calculated Ricci scalar (which identically vanishes since the teleparallel connection is curvature-less), $e=\text{det}\left(\udt{e}{A}{\mu}\right)=\sqrt{-g}$ is the tetrad determinant. The total divergence term defined the boundary quantity $B$ through
\begin{equation}
    \lc{R}=-T+\frac{2}{e}\partial_{\mu}\left(e\udut{T}{\lambda}{\lambda}{\mu}\right):=-T+B\,.
\end{equation}
The boundary term nature of $B$ guarantees that an action formed by a linear contribution of $T$ will produce a teleparallel 
3
 equivalent of general relativity (TEGR) \cite{Hehl:1994ue,Aldrovandi:2013wha}, while modifications of this may produce novel constructions of gravity \cite{Cai:2015emx,Bahamonde:2021gfp}.

The equivalence principle offers a procedure by which to relate local Minkowski frames and the general manifold in GR. TG is not dissimilar in that the gravitational sector is indeed formed by the teleparallel connection but its interaction with matter preserves the minimal coupling prescription, namely \cite{Aldrovandi:2013wha,BeltranJimenez:2020sih}
\begin{equation}
    \partial_{\mu} \rightarrow \mathring{\nabla}_{\mu}\,,
\end{equation}
where partial derivatives are raised to the regular Levi-Civita covariant derivative for matter fields. Thus, both the gravitational and scalar field sectors are developed enough to consider the recently proposed teleparallel analog of Horndeski gravity \cite{Bahamonde:2019shr,Bahamonde:2019ipm,Bahamonde:2020cfv}, also called Bahamonde-Dialektopoulos-Levi Said (BDLS) theory. This construction of gravity in this framework depends on three limiting conditions which arise due to the organically lower-order nature of TG, namely (i) the field equations must be at most second order in their derivatives of the tetrads; (ii) the scalar invariants will not be parity violating; and (iii) the number of contractions with the torsion tensor is limited to being at most quadratic. Without these three conditions, the theory will admit an infinite number of terms which may not contribute appreciably to the physics. Automatically, this implies a weaker form of the generalized Lovelock theorem \cite{Lovelock:1971yv,Gonzalez:2015sha,Gonzalez:2019tky}.

The teleparallel analog of Horndeski gravity or BDLS conditions leads directly to a finite number of contributing scalar invariants, which give the linearly coupled with the scalar field term \cite{Bahamonde:2019shr}
\begin{equation}
    I_2 = v^{\mu} \phi_{;\mu}\,,\label{eq:lin_contrac_scalar}
\end{equation}
where $\phi$ is the scalar field, and while for the quadratic scenario, we find
\begin{align}
J_{1} & =a^{\mu}a^{\nu}\phi_{;\mu}\phi_{;\nu}\,,\label{eq:quad_contrac_scal1}\\
J_{3} & =v_{\sigma}t^{\sigma\mu\nu}\phi_{;\mu}\phi_{;\nu}\,,\\
J_{5} & =t^{\sigma\mu\nu}\dudt{t}{\sigma}{\alpha}{\nu}\phi_{;\mu}\phi_{;\alpha}\,,\\
J_{6} & =t^{\sigma\mu\nu}\dut{t}{\sigma}{\alpha\beta}\phi_{;\mu}\phi_{;\nu}\phi_{;\alpha}\phi_{;\beta}\,,\\
J_{8} & =t^{\sigma\mu\nu}\dut{t}{\sigma\mu}{\alpha}\phi_{;\nu}\phi_{;\alpha}\,,\\
J_{10} & =\udt{\epsilon}{\mu}{\nu\sigma\rho}a^{\nu}t^{\alpha\rho\sigma}\phi_{;\mu}\phi_{;\alpha}\,,\label{eq:quad_contrac_scal10}
\end{align}
where semicolons represent covariant derivatives with respect to the Levi-Civita connection. In addition the regular Horndeski Lagrangian terms (which can be calculating using the regular Levi-Civita connection due to the minimum coupling prescription) \cite{Horndeski:1974wa}
\begin{align}
\mathcal{L}_{2} & :=G_{2}(\phi,X)\,,\label{eq:LagrHorn1}\\
\mathcal{L}_{3} & :=G_{3}(\phi,X)\mathring{\Box}\phi\,,\\
\mathcal{L}_{4} & :=G_{4}(\phi,X)\left(-T+B\right)+G_{4,X}(\phi,X)\left(\left(\mathring{\Box}\phi\right)^{2}-\phi_{;\mu\nu}\phi^{;\mu\nu}\right)\,,\\
\mathcal{L}_{5} & :=G_{5}(\phi,X)\mathring{G}_{\mu\nu}\phi^{;\mu\nu}-\frac{1}{6}G_{5,X}(\phi,X)\left(\left(\mathring{\Box}\phi\right)^{3}+2\dut{\phi}{;\mu}{\nu}\dut{\phi}{;\nu}{\alpha}\dut{\phi}{;\alpha}{\mu}-3\phi_{;\mu\nu}\phi^{;\mu\nu}\,\mathring{\Box}\phi\right)\,,\label{eq:LagrHorn5}
\end{align}
we also arrive at the additional term \cite{Bahamonde:2019shr}
\begin{equation}
    \mathcal{L}_{\text{{\rm Tele}}}:= G_{\text{{\rm Tele}}}\left(\phi,X,T,T_{\text{ax}},T_{\text{vec}},I_2,J_1,J_3,J_5,J_6,J_8,J_{10}\right)\,,
\end{equation}
where the kinetic term is defined as $X:=-\frac{1}{2}\partial^{\mu}\phi\partial_{\mu}\phi$, and where the theory action is represented by
\begin{equation}\label{action}
    \mathcal{S}_{\text{BDLS}} = \frac{1}{2\kappa^2}\int d^4 x\, e\mathcal{L}_{\text{{\rm Tele}}} + \frac{1}{2\kappa^2}\sum_{i=2}^{5} \int d^4 x\, e\mathcal{L}_i+ \int d^4x \, e\mathcal{L}_{\rm m}\,,
\end{equation}
with $\mathcal{L}_{\rm m}$ is the matter Lagrangian in the Jordan conformal frame, $\kappa^2:=8\pi G$, $\lc{G}_{\mu\nu}$ is the standard Einstein tensor, and comma represents regular partial derivatives. We denote the gravitational Lagrangian as $\mathcal{L} = e\left(\mathcal{L}_{{\rm Tele}} + \sum_{i=2}^{5} \mathcal{L}_i\right)$. One small difference to the original form of Horndeski gravity is that the tetrad is being used to make the calculations rather than the metric, but the resulting terms will be identical for the $\mathcal{L}_{2} - \mathcal{L}_{5}$ contributions. Clearly, the standard form of Horndeski gravity is recovered for the limit where $G_{\text{{\rm Tele}}} = 0$. Another important point is that due to the invariance under local Lorentz transformations and diffeomorphisms, BDLS theory will also be invariant under these transformations.

\section{Noether Symmetries}\label{sec:Noeth_symm}

Cosmological models identified through Noether symmetries offer an interesting approach by which to produce candidate models for further investigation. In this section, we present the teleparallel Horndeski Lagrangian followed by a brief review of the Noether approach to producing such cosmological solutions.

\subsection{The point-like Lagrangian}\label{sec:pnt_lk_L}

Here, we will construct the point-like Lagrangian for the Lagrangian in Eq.~\eqref{action} given a spatially flat Friedmann–Lema\^{i}tre–Robertson–Walker (FLRW) metric. To this end, our core task is to incorporate the new torsion scalars into the formulation of the Lagrangian. This generalizes previous works on the topic \cite{Capozziello:2018gms} rendering a much more general set of solutions. Indeed, as will be explored in the Sec.~\eqref{sec:models} this will pose a problem in terms of the presentation of these solutions.

Consider the spatially flat FLRW metric described by
\begin{equation}
    \mathrm{d}s^2 = -  N(t)^2 \mathrm{d}t^2 + a(t)^2 ( \mathrm{d}x^2 +\mathrm{d}y^2+\mathrm{d}z^2)\,,\label{metric_flrw}
\end{equation}
where $N(t)$ represents the lapse function and $a(t)$ the scale factor. This represents the maximally symmetric metric for this spacetime and thus can be used to produce the ensuing equations of motion. Aside from this, the scalar field $\phi$ inherits all the symmetries of the spacetime, namely the position-independence $\phi=\phi\;(t)$.

The metric in Eq.~\eqref{metric_flrw} can be reproduced by the tetrad $\udt{e}{A}{\mu} = \textrm{diag}(N(t),a(t),a(t),a(t))$ which turns out to be compatible with the Weitzen\"{o}ck gauge \cite{Bahamonde:2021gfp}, meaning a vanishing spin connection. Now, we consider the scalar contributions to the point-like Lagrangian by first calculating the torsion scalar
\begin{equation}
    T = \frac{6\dot{a}^2}{a^2N^2}\,,
\end{equation}
which is the Lagrangian density for TEGR. The only linear contraction scalar \eqref{eq:lin_contrac_scalar} is then given by
\begin{equation}
    I_2=\frac{3\dot{a}\dot{\phi}}{aN^2}\,.
\end{equation}
Interestingly, all the quadratic contraction scalars (\ref{eq:quad_contrac_scal1}--\ref{eq:quad_contrac_scal10}) vanish since only the vector irreducible is nonzero for this tetrad.

For completeness we also show the other scalars that are used to form the point-like Lagrangian where the derivative operators on the scalar field produce
\begin{align}
    \lc{\Box}{\phi} &= -\frac{\ddot{\phi}}{N^2}-\frac{3\dot{a}\dot{\phi}}{a N^2}+\frac{\dot{\phi}\dot{N}}{N^3}\,,\\
    (\lc{\nabla}_{\mu}\lc{\nabla}_{\nu}\phi)^2 &= \frac{3\dot{a}^2\dot{\phi}^2}{a^2N^4}+\frac{1}{N^4}\left(\ddot{\phi}-\dfrac{\dot{\phi}\dot{N}}{N}\right)^2\,,\\
    (\lc{\nabla}_{\mu}\lc{\nabla}_{\nu}{\phi})^3 &= -\frac{3\dot{a}^3\dot{\phi}^3}{a^3N^6}-\frac{1}{N^6}\left(\ddot{\phi}-\dfrac{\dot{\phi}\dot{N}}{N}\right)^3\,,
\end{align}
while the kinetic term and the Ricci scalar are given by
\begin{equation}
X = \dfrac{\dot{\phi}^2}{2N^2}\,, \quad \mathring{R} = -T+B = \dfrac{6\ddot{a}}{aN^2}+\dfrac{6\dot{a}^2}{a^2N^2}-\dfrac{6\dot{a}\dot{N}}{aN^3}\,.
\end{equation}

Additionally, we apply the Lagrangian multiplier approach to the point-like Lagrangian by introducing the multipliers $\lambda_1$ and $\lambda_2$ which correspond to the scalars $T$ and $I_2$ respectively. The reasoning behind the use of Lagrangian multipliers is to identify any possible problematic features that may appear in the behaviour of the model. Moreover, this will provide a more convenient approach to investigating the properties of the theory more effectively.

By incorporating the Lagrangian multipliers as well as the scalar substitutions into the teleparallel Horndeski Lagrangian, we determine the point-like Lagrangian to be
\begin{align}
    \mathcal{L}=&a^3N\left(G_2(\phi,X)+G_{{\rm Tele}}(\phi,X,T,I_2)+ T G_4(\phi,X)-I_2 G_{ {\rm Tele},I2}(\phi,X,T,I2)-T G_{ {\rm Tele},T}(\phi,X,T,I2)\right)+\nonumber \\
    &
    +\frac{6a\ddot{a}}{N^3}\left(a N^2 G_4(\phi,X) + G_5(\phi) \dot{a}\dot{\phi}\right) +\frac{a\ddot{\phi}}{N^3}\left(a^2N^2G_3(\phi,X)+3G_5(\phi)\dot{a}^2+6 a G_{4,X}(\phi,X)\dot{a}\dot{\phi}\right)-\nonumber \\
    &- \frac{a^2\dot{N}}{N^2}\left(6 G_4(\phi,X)\dot{a}+aG_3(\phi,X)\dot{\phi}\right) - \frac{3a\dot{a}\dot{N}\dot{\phi}}{N^4}\left(3 G_5(\phi,X)\dot{a}+2a G_{4,X}(\phi,X)\dot{\phi}\right) + \nonumber \\
    &+ \frac{3a\dot{a}}{N}\left(a G_3(\phi,X)\dot{\phi}+a G_{{\rm Tele},I2}(\phi,X,T,I_2)\dot{\phi}+2 G_{{\rm Tele},T}(\phi,X,T,I_2)\dot{a}\right) + \nonumber \\ 
    &+ \frac{3a\dot{a}^2\dot{\phi}}{N^3}\left(G_5(\phi,X)\dot{a}+2a G_{4,X}(\phi,X)\dot{\phi}\right) \,.
\end{align}

For this setting, the Lagrangian coordinates are $a(t),\phi(t),N(t),T(t),I_2(t)$ which implies a configuration space setup $Q=(a,\phi,N,T,I_2)$. Another feature of this Lagrangian is that the second-order derivatives can readily be eliminated through integration by parts. On the other hand, the only problematic term in this process is the $a^3\ddot{\phi}\;G_3/N$ contribution which fails to be taken away using this approach. In order to resolve this issue, a reasonable choice is made in which we set
\begin{equation}
    G_{3XX}=0 \Rightarrow G_3(\phi,X)=g(\phi)X+h(\phi)\,.\label{eq:G3_redef}
\end{equation}
While not physically motivated, the choice that $G_{3XX}=0$ provides a route by which most of the physically interesting and well-posed formulations can be derived. On this point, observational data cannot yet differentiate between the plethora of models available, and so we pursue the set of models that result from this setting providing a new range of possible cosmological models to probe.

Hence, the point-like Lagrangian in this setting is given by 
\begin{align}
    \mathcal{L} = &a^3 N \left(G_2(\phi ,X)+T G_4(\phi ,X)-I_2 G_{{\rm Tele},I2}(\phi ,X,T,I_2)-T G_{\rm Tele,T}(\phi ,X,T,I_2)+G_{{\rm Tele}}(\phi ,X,T,I_2)\right)-\nonumber \\
    &-\frac{3 a \dot{a}^2}{N^3}\left(\dot{\phi}^2 \left(G_5'(\phi )-2 G_{4,X}(\phi ,X)\right)+2 N^2 \left(2 G_4(\phi ,X)-G_{{\rm Tele},T}(\phi ,X,T,I_2)\right)\right)+\nonumber \\
    &+\frac{a^2 \dot{a} \dot{\phi}}{N^3}\left(3 N^2 \left(G_{ {\rm Tele},I2}(\phi ,X,T,I_2)-2 G_{4,\phi}(\phi ,X)\right)+g(\phi ) \dot{\phi}^2\right)-\frac{a^3 \dot{\phi}^4 g'(\phi )}{6 N^3}-\frac{a^3 \dot{\phi}^2 h'(\phi )}{N}\,,\label{eq:Lagran_min}
\end{align}
which is the minimal form of the teleparallel Horndeski in this setting.

\subsection{Noether Symmetries}\label{sec:noether_symm_app}

We briefly review how a general differential equation behaves under the action of a point transformation. Consider a system governed by a Lagrangian $\mathcal{L}$ with $n$ generalized coordinates $q^i$ and an independent variable $t$. The general form of an infinitesimal transformation acting on that system is expressed as follows

Suppose that the dynamics of a system are governed by a Lagrangian $\mathcal{L}$ in terms of n generalized coordinates $q^i$ while $t$ is the independent variable.
\begin{equation}
    t\Rightarrow t'=t+\epsilon\;\xi(q^i,t)\;\;, \;\;q^i\Rightarrow q^{i'}=q^i+\epsilon \eta^i(q^i,t)\,,
\end{equation}
which can be encapsulated in the generator vector of the transformation
\begin{equation}
    \mathcal{X}=\xi(q^i,t)\dfrac{\partial}{\partial t}\;+ \eta^i(q^i,t)\dfrac{\partial}{\partial q^i}\,.
\end{equation}
For any differentiable function $F$, the action of this transformation is given by
\begin{equation}
    F(q',t')=F(q,t)+ \epsilon\mathcal{X}(F(q,t)) +O(\epsilon^2)\,,
\end{equation}
which can readily be extended for the case where $F$ also has a velocity dependence
\begin{equation}
    F(q',\dot{q}',t')=F(q,\dot{q},t)+\epsilon\mathcal{X}^{(1)}(F(q,\dot{q},t))+O(\epsilon^2)\,,
\end{equation}
where
\begin{equation}
    \mathcal{X}^{(1)}=\mathcal{X}+(\dot{\eta}^i-\dot{q}^i\dot{\xi}\;)\dfrac{\partial}{\partial \dot{q}^i}\,,
\end{equation}
is the first prolongation of the generator vector. For the case in which $F = F(t,q,\dot{q},\ddot{q})$, the second prolongation will be applied giving
\begin{equation}
    \mathcal{X}^{(2)}=\mathcal{X}^{(1)}+(\ddot{\eta}^i-\dot{q}^i\ddot{\xi}-2\ddot{q}^i\dot{\xi}\;)\dfrac{\partial}{\partial \dot{q}^i}\,,
\end{equation}
and so on for the $n$th order time derivative of $q$.

For the system under study consider the Lagrangian $\mathcal{L} = \mathcal{L}(t,q,\dot{q})$, the action of the system is said to invariant under infinitesimal transformations (up to a total divergence term), if the Rund-Trautman identity holds, namely \cite{Basilakos:2011rx}
\begin{equation}
    \mathcal{X}^{(1)} \;\mathcal{L} + \dfrac{\mathrm{d}{\xi}}{\mathrm{d}t} \;\mathcal{L}\;= \;\dfrac{\mathrm{d}f}{\mathrm{d}t}\,,\label{eq:Rund-Trautman}
\end{equation}
where $\mathcal{X}^{(1)}$ is the first prolongation of the generating vector. Then, the generator vector $\mathcal{X}$ is a Noether symmetry of the dynamical system described by $\mathcal{L}$. For any such Noether symmetry, there exists a function
\begin{equation}
    I(t,q,\dot{q}\;)= f -\mathcal{L}\;\xi-\dfrac{\partial\mathcal{L}}{\partial \dot{q}^i}(\eta^i-\dot{q}^i\xi)\,,
\end{equation}
which is a first integral of the equations of motion. In the following, we will give explicit examples of the above general scenario.

\section{Classification of teleparallel Horndeski Cosmologies}\label{sec:models}

For our configuration space, namely $Q=(a,\phi,N,T,I_2)$, of the point-like Lagrangian in Eq.~\eqref{eq:Lagran_min}, together with the independent variable of cosmic time $t$, the generator vector is described by
\begin{equation}
    \mathcal{X}=\xi(t,a,\phi,N,T,I_2)\;\partial_t+\mathit{\Sigma } \eta_{q_i}(t,a,\phi,N,T,I_2)\;\partial q_i\;,\;\;q_i=a,\phi,N,T,I_2\,.
\end{equation}
By applying the Rund-Trautman identity in Eq.~\eqref{eq:Rund-Trautman} to the point-like Lagrangian \eqref{eq:Lagran_min} produces 62 differential equations for the coefficients of the Noether vector $\xi,\eta_a,\eta_{\phi},\eta_N,\eta_T,\eta_{I_2},f$ and the arbitrary Lagrangian functions $G_i(\phi,X)$. As already discussed, the Lagrangian functions are not all independent of each other \cite{Bahamonde:2019shr,Bahamonde:2019ipm,Bahamonde:2020cfv} and may feature some overlap.

Considerations from Noether symmetries alone are not necessarily enough to fully determine $G_i(\phi,X)$ models. In some works, specific forms of the $G_i(\phi,X)$ functions are assumed in standard Horndeski theories \cite{Dimakis:2017zdu,Paliathanasis:2014rja,Dimakis:2017kwx,Giacomini:2017yuk} which aides in the full determination of cosmological models, as well as the Noether vector coefficients. Our strategy is rather to consider the most general Lagrangian and to constrain as much as possible the unknown functions of the model Lagrangian and Noether vector coefficients. This involves treating the symmetries in the most general way possible and investigating each possible symmetry in turn. This will produce particular models for the various scenarios that are produced in the teleparallel analogue of Horndeski gravity. In our case, if at least one coefficient of the generating Noether vector is nonzero, then a Noether symmetry is said to exist. The existence of such symmetries leads to different forms of the $G_i(\phi,X)$ functions which may be physically interesting.

In this work we exhaustively explored every possible cases that arise from considering the Noether symmetries as applied to the teleparallel analogue of Horndeski gravity. These symmetries are determined by the system of equations that come about by using the point-like Lagrangian \eqref{eq:Lagran_min} in conjunction with the Noether condition in Eq.~\eqref{eq:Rund-Trautman} which leads to the over-constrained system of 62 equations. Now, these solutions are impacted by whether the $g(\phi)$ function is vanishing or not in the redefinition in Eq.~\eqref{eq:G3_redef}, this leads to genuinely distinct solutions. Moreover, the Noether classification cases result by considering the vanishes or not of each Noether vector coefficient, which lead to distinct solutions in most cases. The enormity of the teleparallel analogue of Horndeski gravity means that this process will result in many cases some of which may involve lengthy solutions not appropriate for such a setting. For this reason, we show below four specific examples of these cases, and retain the full set of classification case solutions separately\footnote{The full set of Noether symmetry solutions can be found at \href{https://github.com/jacksonsaid/BDLS_Noether_classification.git}{https://github.com/jacksonsaid/BDLS\_Noether\_classification.git}}.\\

\noindent \underline{Case 1 (2.a.ii.1.a.i.1.b.i.1.b in Table 2a):} \\

In this first example, we find a solution to the 62 differential equations in which the Noether coefficients turn out to be
\begin{align}
    \xi (t,a,\phi,N,T,I2) = &\tilde{\xi} (t)\,, \\
    \eta_a (t,a,\phi,N,T,I2) = &-\frac{1}{3} c_1 a\,, \\
    \eta_{\phi} (t,a,\phi,N,T,I2) = &\frac{c_1 g(\phi)}{g'(\phi)}\,, \\
    \eta_N (t,a,\phi,N,T,I2) = &N \left(c_1 - \xi (t) - \frac{c_1 g(\phi) g''(\phi)}{g'(\phi)^2} \right)\,, \\
    \eta_T (t,a,\phi,N,T,I2) = &2 c_1 T \left( \frac{g(\phi) g''(\phi)}{g'(\phi)^2}-1\right)\,, \\
    \eta_{I_2} (t,a,\phi,N,T,I2) = &c_1 \Bigg(I_2 \left( - \frac{c_1 g(\phi) g''(\phi)}{g'(\phi)^2}-1\right) + \frac{4}{(2c_2+3)g(\phi)^3}\Big(g(\phi) g'(\phi)\tilde{G}_4'(\phi)\nonumber\\
    &+ \tilde{G}_4(\phi) \left( g(\phi)g''(\phi)-2 g'(\phi)^2\right) \Big) \Bigg)\,, \\
    f (t,a,\phi,N,T,I2) =  &c_7\,,
\end{align}
while the BDLS model functions ae given by
\begin{align}
    G_2(\phi,X) = &\frac{g'(\phi)}{2}\left(2 c_3 + c_3 X + 2 X^2\right)- \tilde{G}_{{\rm Tele}}(\phi,X)\,,\\
    G_3(\phi,X) = &c_4 + g(\phi) \left(c_5+X\right)\,,\\
    G_4(\phi,X) = &\tilde{G}_4(\phi) + \frac{(2c_2 +3)X g(\phi)^2}{4 g'(\phi)}\,,\\
    G_5(\phi,X) = &c_6 + \int _1 ^{\phi} \frac{c_2 g(x)^2}{g'(x)}dx\,,\\    
    G_{{\rm Tele}}(\phi,X,T,I_2) = &\tilde{G}_{{\rm Tele}}(\phi,X) +\frac{1}{4}\Big( 4 \tilde{G}_4(\phi) T + 2 g(\phi) I_2 (c_3 - 4c_5 +2X (3+2c_2))+4 \bar{G}_{{\rm Tele}}(\frac{g(\phi)^2 T}{g'(\phi)^2})g'(\phi)+\nonumber \\
    &+8 I_2 \tilde{G}_4'(\phi) +\frac{(2c_2+3)g(\phi)^3 I_2 T}{g'(\phi)^2} + \frac{2g(\phi)^2 X (2c_2 T g'(\phi)-(2c_2+3)I_2g''(\phi))}{g'(\phi)^2}\Big)\,.   
\end{align}
which satisfies, as all the cases in the work, the speed constraint of gravitation waves \cite{Bahamonde:2019ipm}. In this and the other classification cases, the denominators cannot vanish due to the particular case not allowing it. This renders each case safe from divergences. This is an interesting case where the $G_4$ functional expresses some dependence on the kinetic while the $G_5$ function is nonzero. This is balanced by an intricate form of $G_{ {\rm Tele}}$ which now must contain several terms to satisfy the gravitational wave constraint condition. It is difficult to relate these models to their standard Horndeski analogue since they do not regularly allow for so much of the dynamics to feature in the $G_5$ function. On the other hand, we do note the dependence $G_5 = G_5 (\phi)$ which omits the kinetic term, while the other functions do feature this term. \\

\noindent \underline{Case 2 (2.a.ii.1.a.ii.2.a.ii in Table 2a):}\\

In other case of the solutions of the 2 differential equations, we find the Noether coefficients have values
\begin{align}
    \xi (t,a,\phi,N,T,I2) = &\tilde{\xi} (t)\,, \\
    \eta_a (t,a,\phi,N,T,I2) = &\frac{3 c_1 - a^3 \tilde{\eta}_{\phi}(t,a,\phi,N,I_2) g'(\phi)}{3 a^2 g(\phi)}\,, \\
    \eta_{\phi} (t,a,\phi,N,T,I2) = &\tilde{\eta}_{\phi}(t,a,\phi,N,I2)\,, \\
    \eta_N (t,a,\phi,N,T,I2) = &N \left( - \frac{3 c_1}{a^3 g(\phi)}+ \frac{\tilde{\eta}_{\phi}(t,a,\phi,N,I2)g'(\phi)}{g(\phi)} - \tilde{\xi}(t) - \frac{\tilde{\eta}_{\phi}(t,a,\phi,N,I2)g''(\phi)}{g'(\phi)}\right)\,, \\
    \eta_T (t,a,\phi,N,T,I2) = &\frac{1}{c_2 a^3 N g(\phi)^2 g'(\phi)}\Big(f'(t) g'(\phi)^2 +a^3 N \tilde{\eta}_{\phi}(t,a,\phi,N,I2)\Big(-2 c_2 T g(\phi)g'(\phi)^2 +\nonumber \\
    &+ 2 c_2 T g(\phi)^2 g''(\phi) - g'(\phi)^2 \tilde{G}_{{\rm Tele}}'(\phi)+ g'(\phi)g''(\phi)\tilde{G}_{ {\rm Tele}}(\phi)\Big)\Big)\,, \\
    \eta_{I_2} (t,a,\phi,N,T,I2) = &\eta_{I_2} (t,a,\phi,N,T,I2)\,, \\
    f (t,a,\phi,N,T,I2) =  &f(t)\,,
\end{align}
where some of these results remain very general, and where the BDLS model functions now take on the form
\begin{align}
    G_2(\phi,X) = &G_2(\phi,X)\,,\\
    G_3(\phi,X) = &c_3 + g(\phi) \left(c_4+X\right)\,,\\
    G_4(\phi,X) = &\frac{c_2 g(\phi)^2}{g'(\phi)}\,,\\
    G_5(\phi,X) = &c_5\,, \\    
    G_{ {\rm Tele}}(\phi,X,T,I_2) = &- G_2(\phi,X) + \tilde{G}_{{\rm Tele}}(\phi) +\frac{4}{3} c_2  I_2 g(\phi) + 2 c_4 X g'(\phi) + X^2 g'(\phi) -\nonumber \\
    &- \frac{8}{3} c_2 X g'(\phi) - \frac{2 c_2 g(\phi)^2 I_2 g''(\phi)}{g'(\phi)^2}\,.
\end{align}
This case is different because the $G_5$ function is fully determined to be a constant while the $G_4$ functional does not depend on the kinetic term. To that end, these parts of the model would survive the gravitational wave speed constraint in standard Horndeski gravity. The BDLS correction term then adds news dynamics distinct to this standard Horndeski model. What is interesting in this case is that the $G_{ {\rm Tele}}$ term inherits directly the functional from the $G_2$ term producing a coupling between the basic standard term and the BDLS correction term. \\

\noindent \underline{Case 3 (2.b.i2.b.ii.2.b.ii.2 in Table 2b):}\\

In a similar vein, the generality of the $G_2$ functional and the independence of $G_4$ from the kinetic term remains the case with this case in which the Noether coefficients assume the values
\begin{align}
    \xi (t,a,\phi,N,T,I2) = &\tilde{\xi} (t)\,, \\
    \eta_a (t,a,\phi,N,T,I2) = &\frac{c_1}{c_2 a^2}\,, \\
    \eta_{\phi} (t,a,\phi,N,T,I2) = &0\,, \\
    \eta_N (t,a,\phi,N,T,I2) = &- N \tilde{\xi}(t)\,, \\
    \eta_T (t,a,\phi,N,T,I2) = &- \frac{3 c_1 (2 \tilde{G}_{{\rm Tele}}(\phi)+T G_5'(\phi))}{c_2 a^3 G_5'(\phi)}\,,\\
    \eta_{I_2} (t,a,\phi,N,T,I2) = &\eta_{I_2} (t,a,\phi,N,T,I2)\,, \\
    f (t,a,\phi,N,T,I2) =  &c_3\,,
\end{align}
while the Horndeski functions take the form
\begin{align}
    G_2(\phi,X) = &G_2(\phi,X)\,,\\
    G_3(\phi,X) = &c_4 + c_5 X\,,\\
    G_4(\phi,X) = &\tilde{G}_4(\phi) +\frac{X}{2}G_5'(\phi)\,,\\
    G_5(\phi,X) = &G_5(\phi)\,, \\    
    G_{{\rm Tele}}(\phi,X,T,I_2) = &-G_2(\phi,X) + \frac{(\tilde{G}_{ {\rm Tele}}(\phi)+TG_5'(\phi))(2 \tilde{G}_4(\phi)+X G_5'(\phi))}{G_5'(\phi)} +\nonumber \\
    &+ I_2 \left(c_6+2\tilde{G}_4'(\phi) + X(G_5''(\phi)-c_2) \right)\,.
\end{align}
This case generalizes the $G_5$ functional to arbitrary dependence on the scalar field, while also putting some kinetic term dependence on the $G_4$ term. This is an example of a model that satisfies the gravitational wave constraint for BDLS theory but not for standard Horndeski gravity. As in the previous two cases, there is an $I_2$ dependence on $G_{ {\rm Tele}}$, together with a coupling with both $G_4$and $G_5$ functionals. \\

\noindent \underline{Case 4 (2.b.ii1.a.ii.2.b.i in Table 2b):}\\

The final case we consider that solves the 62 differential equations of the Noether symmetries gives
\begin{align}
    \xi (t,a,\phi,N,T,I2) = &\tilde{\xi} (t)\,, \\
    \eta_a (t,a,\phi,N,T,I2) = &\frac{c_1}{\sqrt{a}}\,, \\
    \eta_{\phi} (t,a,\phi,N,T,I2) = &0\,, \\
    \eta_N (t,a,\phi,N,T,I2) = &- N \tilde{\xi}(t)\,, \\
    \eta_T (t,a,\phi,N,T,I2) = &- \frac{3 c_1 T}{a^{3/2}}\,,\\
    \eta_{I_2} (t,a,\phi,N,T,I2) = &- \frac{3 c_1 I_2}{2a^{3/2}}\,, \\
    f (t,a,\phi,N,T,I2) =  &c_2\,,
\end{align}
where the model functionals take the form
\begin{align}
    G_2(\phi,X) = &\tilde{G}_2(\phi) - \tilde{G}_{ {\rm Tele}}(\phi,X)\,,\\
    G_3(\phi,X) = &c_3\,,\\
    G_4(\phi,X) = &\bar{G}_4' (\phi) + X \tilde{G}_4 '(\phi)\,,\\
    G_5(\phi,X) = &G_5(\phi)\,, \\
    G_{ {\rm Tele}}(\phi,X,T,I_2) = &-\tilde{G}_2(\phi) + \tilde{G}_{ {\rm Tele}}(\phi,X) + \bar{G}_{{\rm Tele}}(\phi) I_2 \sqrt{T} + \hat{G}_{{\rm Tele}}(\phi) T -\nonumber \\
    &- \tilde{G}_4 (\phi)T X + 2 I_2 (\bar{G}_4' (\phi) + X \tilde{G}_4 '(\phi))+\frac{3XT}{2}G_5'(\phi)\,.
\end{align}
As in case 3, one of the Noether coefficients vanishes making the system slightly easier to solve. Here, we again observe nonvanishing $G_4$ and $G_5$ showing another case which is only possible, in an observationally consistent manner, in BDLS as compared with standard Horndeski theory.

\section{Conclusions}\label{sec:conc}
Horndeski gravity is the most general scalar-tensor framework giving to second-order field equations \cite{Horndeski:1974wa}. Its {\rm Tele}parallel analogue \cite{Bahamonde:2019shr} offers a larger framework in which to construct model since torsion naturally produces lower-order theories of gravity. The Teleparallel of Horndeski gravity \cite{Bahamonde:2019ipm} is especially interesting because provides a direct way to circumvent the recent speed constraint on the propagation of gravitational waves \cite{Ezquiaga:2018btd}. Moreover it is well known that most higher order theories in  curvature-based gravity can be mapped onto a dynamically equivalent Horndeski model, it follows that the Teleparallel analogue allows for even more higher order dynamically equivalent theories. For these reasons, in this work, we studied the symmetries that emerge from the Noether symmetry approach. When such symmetries arise, the equations of motion become reducible and solvable in most cases. This means that exact solutions, which are rare in general, become more obtainable within this approach.

We do this by first determining the point-like Lagrangian for our setting in Sec.~\ref{sec:pnt_lk_L} using the maximally symmetric form of the flat FLRW metric. This is performed using a tetrad that is compatible with the Weitzenb\"{o}ck gauge. We then remove second-order derivatives in this Lagrangian using integration by parts. However, one of the terms, related to $G_3$ poses a problem for this procedure and so we take the still general form of Eq.~\eqref{eq:G3_redef} for its functional form. In the end, this produced the point-like Lagrangian in Eq.~\eqref{eq:Lagran_min} where the second-order derivatives have been removed. We then lay out the general approach taken in the remainder of the work in Sec.~\eqref{sec:noether_symm_app} where the first order equation of motion that emerges out of the Noether symmetry is given.

Thus, by applying the Rund-Trautman identity in Eq.~\eqref{eq:Rund-Trautman} to the point-like Lagrangian \eqref{eq:Lagran_min} we find a system of 62 differential equations for the Noether vector coefficients and model functionals. This in turn leads to a large number of cases of this general system of equations. Given the enormity of these cases we display them elsewhere (\href{https://github.com/jacksonsaid/BDLS_Noether_classification.git}{GitHub}). Saying that, in Sec.~\eqref{sec:models} we showcase 4 of these cases which display the power of this approach. To varying degrees, these cases give a determined system of model functionals and Noether symmetry coefficients. To display some information about the remainder of the other cases we present Tables in App.~\ref{sec:app} where all the solution scenarios are enumerated together with the conditions which defines them. These are divided by whether the $G_{4,XX}$ vanishes or not which is then subdivided into subcases.

These tables show the complexity available through the Teleparallel analogue of Horndeski gravity. Moreover, they provide a wealth of models which may provide interesting dynamics for further investigation. Given the breadth of classification cases, it would not be feasible to explore the cosmology that they instigate in a systematic way. However, it would be very interesting to consider the viable models further and to determine their cosmological evolution.

\begin{acknowledgments}
KFD acknowledges support by the Hellenic Foundation for Research and Innovation (H.F.R.I.) under the “First Call for H.F.R.I. Research Projects to support Faculty members and Researchers and the procurement of high-cost research equipment grant” (Project Number: 2251). The authors would like to acknowledge networking support by the COST Action CA18108 and funding support from Cosmology@MALTA which is supported by the University of Malta. The authors would also like to acknowledge funding from ``The Malta Council for Science and Technology'' in project IPAS-2020-007.
\end{acknowledgments}

\appendix
\section{Noether Classifications}\label{sec:app}

In the appendix we present on a table all the different classes of theories that appear through the classification process. Not all of them lead to symmetries because in some cases the system of equations was too complicated to be solved, or not all the functions are fully determined. In greater detail, meaning the Noether vector coefficients and the form of the $G_i$ functions of the Lagrangian, can be found in the notebook files in \href{https://github.com/jacksonsaid/BDLS_Noether_classification.git}{GitHub}. 

\vspace{0.25cm}

{\mathversion{bold} $ 1. : G_{4,XX}\neq 0$}

\vspace{2mm}

\begin{tabular}{ cc }
\begin{tabular}{ | m{1,6cm} | m{5.2cm}|| } 
  \hline
   
  \multicolumn{2}{|c|}{\mathversion{bold}$1$.\bf{a} :  \mathversion{bold}$ \mathbf{ h'(\phi)\neq 0}$}\\
 \hline
  1.a.i &$ G_{{\rm Tele },I_2 I_2}(\phi,X,T,I_2)\neq0$ \\ 
  \hline
 1.a.ii &$G_{{\rm Tele},I_2I_2} (\phi,X,T,I_2)=0$ \\ 
  \hline
 1.a.ii.1 & $G_{{\rm Tele}2,T}(\phi,X,T)\neq0$ \\
 \hline
  1.a.ii.2 & $G_{{\rm Tele}2,T}(\phi,X,T)=0$  \\ 
  \hline
\end{tabular}&

\vspace{4mm}

\begin{tabular}{ | m{1.6cm} | m{6cm}|| } 
  \hline
  \multicolumn{2}{|c|}{\mathversion{bold}$1$.\bf{b} :  \mathversion{bold}$ \mathbf{ h'(\phi)=0}$}\\
 \hline
 1.b.i &$ G_{{\rm Tele},I_2 I_2}(\phi,X,T,I_2)\neq0$  \\ 
  \hline
  1.b.ii & $G_{{\rm Tele},I_2I_2}(\phi,X,T,I_2) =0$ \\
  \hline
  1.b.ii.1 & $G_{{\rm Tele}2,I_2I_2}(\phi,X,I_2)\neq 0$ \\
  \hline
  1.b.ii.2 & $G_{{\rm Tele}2,I_2I_2}(\phi,X,I_2)=0$  \\ 
  \hline
\end{tabular}\\
\end{tabular}

\vspace{0.25cm}

{\mathversion{bold} $ 2. : G_{4,XX}= 0$}

\vspace{2mm}

%\end{center}

\begin{tabular}{ | m{3cm} | m{13cm}|| } 
  \hline
   
  \multicolumn{2}{|c|}{\mathversion{bold}$2$.\bf{a} :  \mathversion{bold}$ \mathbf{G_{{\rm Tele},{XI2}}(\phi,X,T,I_2)\neq2 G_{4b}'(\phi)-g(\phi)}$}\\
 \hline
  
  2.a.i. &$ G_{{\rm Tele},{XT}}(\phi,X,T,I_2)\neq G_5'(\phi)$ \\ 
  \hline
 2.a.i.1.&$g(\phi)\neq0$  \\ 
  \hline
  2.a.i.1.a. & $G_{{\rm Tele} 2a,{TT}}(\phi,T)\neq-G_{{\rm Tele}1,{TT}} (\phi,T,I2)$ \\ 
  \hline
  2.a.i.1.b. & $G_{{\rm Tele} 2a,{TT}}(\phi,T)\,=\,-G_{{\rm Tele}1,{TT}} (\phi,T,I2)$\\ 
  \hline
  2.a.i.1.b.i. & $  G_{{\rm Tele} 1b,{I_2}}(\phi,I_2) \neq0$ \\ 
  \hline
  2.a.i.1.b.ii. & $  G_{{\rm Tele} 1b,{I_2}}(\phi,I_2)=0$ \\
  \hline
  2.a.i.1.b.ii.1. & $G_{{\rm Tele} 3{T I_2}}(\phi,I,I_2),\neq0$ \\
  \hline
  2.a.i.1.b.ii.2. &$G_{{\rm Tele} 3,{T I_2}}(\phi,I,I_2)=0$\\
  \hline
   2.a.i.1.b.ii.2.a. & $G_{{\rm Tele} 2c,{TT} }(\phi,T) \,+ \,G_{{\rm Tele} 3a,{TT}}(\phi,T)\neq0$\\
  \hline
  2.a.i.1.b.ii.2.b. &$G_{{\rm Tele} 2c,{TT} }(\phi,T)\,+ \,G_{{\rm Tele} 3a,{TT}}(\phi,T)=0$\\
  \hline\hline
    2.a.i.2.& $g(\phi)=0$  \\ 
  \hline
   2.a.i.2.a.& $G_{4b}(\phi)\neq G_5'(\phi)/2$  \\ 
  \hline
   2.a.i.2.a.i.& $G_{{\rm Tele} 1,{I_2}}(\phi,T,I_2)\neq 2 G_{4b}'(\phi)$  \\ 
  \hline
  2.a.i.2.a.i.1. & $h'(\phi)\neq 0$  \\ 
  \hline
  2.a.i.2.a.i.1.a. & $G_{{\rm Tele} 3,{I_2I_2}}(\phi,T,I_2)\neq0$  \\ 
  \hline
     2.a.i.2.a.i.1.b. & $G_{{\rm Tele} 3,{I_2I_2}}(\phi,T,I_2)=0$  \\ 
  \hline
   2.a.i.2.a.i.1.b.i. & $G_{{\rm Tele} 3b,T}(\phi,T)\neq0$  \\ 
  \hline
\end{tabular}

\begin{tabular}{ | m{3cm} | m{13cm}|| } 
  \hline
     2.a.i.2.a.i.1.b.ii. & $G_{{\rm Tele} 3b,T}(\phi,T)=0$  \\ 
  \hline
   2.a.i.2.a.i.1.b.ii.1. & $G_{{\rm Tele} 1,{TI_2}}(\phi,T,I_2)\neq0$  \\ 
  \hline
   2.a.i.2.a.i.1.b.ii.2. & $G_{{\rm Tele} 1,{TI_2}}(\phi,T,I_2)=0$  \\ 
  \hline
  2.a.i.2.a.i.1.b.ii.2.a & $G_{{\rm Tele}1,{I_2I_2}}(\phi,I_2)\neq0$  \\ 
  \hline
  2.a.i.2.a.i.1.b.ii.2.b & $G_{{\rm Tele}1,{I_2I_2}}(\phi,I_2)=0$  \\ 
  \hline
  2.a.i.2.a.i.2. & $ h'(\phi)=0$  \\ 
  \hline
   2.a.i.2.a.ii. & $G_{{\rm Tele} 1,{I_2}}(\phi,T,I_2)=2 G_{4b}'(\phi)$  \\ 
  \hline
  2.a.i.2.a.ii.1. & $G_{4b}(\phi)\neq c_2/h'(\phi)+G_5'(\phi)/2$  \\ 
  \hline
   2.a.i.2.a.ii.1.a. & $c_3\neq0$  \\ 
  \hline
   2.a.i.2.a.ii.1.b. & $c_3=0$  \\ 
  \hline
  2.a.i.2.a.ii.1.b.i & $G_{{\rm Tele} 3,{I_2I_2}}(\phi,T,I_2)\neq0$  \\ 
  \hline
  2.a.i.2.a.ii.1.b.ii & $G_{{\rm Tele} 3,{I_2I_2}}(\phi,T,I_2)=0$  \\ 
  \hline
  2.a.i.2.a.ii.2. & $G_{4b}(\phi)= c_2/G_{3h}'(\phi)+G_5'(\phi)/2$  \\ 
  \hline
  2.a.i.2.a.ii.2.a & $G_{{\rm Tele} 3,{I_2I_2}}(\phi,T,I_2)\neq0$  \\ 
  \hline
  2.a.i.2.a.ii.2.b & $G_{{\rm Tele} 3,{I_2I_2}}(\phi,T,I_2)=0$  \\ 
  \hline
  2.a.i.2.a.ii.2.b.i & $G_{{\rm Tele} 3b,{T}}(\phi,T)\neq0$  \\ 
  \hline
  2.a.i.2.a.ii.2.b.ii & $G_{{\rm Tele} 3b,{T}}(\phi,T)=0$  \\ 
  \hline
  2.a.i.2.b. & $G_{4b}(\phi)= G_5'(\phi)/2$  \\ 
  \hline
  2.a.i.2.b.i & $h'(\phi)\neq 0$  \\ 
  \hline
  2.a.i.2.b.ii & $h'(\phi)= 0$  \\ 
  \hline\hline
   2.a.ii. & $G_{{\rm Tele},{XT}}(\phi,X,T,I_2)= G_5'(\phi)$  \\ 
  \hline
  2.a.ii.1.& $g(\phi)\neq0$  \\ 
  \hline
  2.a.ii.1.a& $G_{4b}(\phi)\neq G_5'(\phi)/2$  \\ 
  \hline
  2.a.ii.1.a.i & $G_{{\rm Tele}2,{TI_2}}(\phi,T,I_2)\neq 0$  \\ 
  \hline
  2.a.ii.1.a.i.1 & $G_5'(\phi)\neq0$  \\ 
  \hline
  2.a.ii.1.a.i.1.a & $G_{{\rm Tele}1,{XI_2}}\neq 2 G_{4b}'(\phi)$  \\ 
  \hline
  2.a.ii.1.a.i.1.a.i & $3 g(\phi)^2 +2g'(\phi)(- 2G_{4b}(\phi)+G_5'(\phi)) +g(\phi) ( -6G_{4b}'(\phi)+ 3G_{{\rm Tele}1,{XI_2}}(\phi,X,I_2))\neq0$  \\ 
  \hline
  2.a.ii.1.a.i.1.a.ii & $3 g(\phi)^2 +2g'(\phi)(- 2G_{4b}(\phi)+G_5'(\phi)) +g(\phi) ( -6G_{4b}'(\phi)+ 3G_{{\rm Tele}1,{XI_2}}(\phi,X,I_2)) = 0$  \\ 
  \hline
  2.a.ii.1.a.i.1.b & $G_{{\rm Tele}1,{XI_2}}=2 G_{4b}'(\phi)$  \\ 
  \hline
\end{tabular}

\newpage

\begin{tabular}{ | m{3cm} | m{14cm}|| } 
  \hline
   
   2.a.ii.1.a.i.1.b.i & $G_{4b}(\phi)\neq 3g(\phi)^2/4g'(\phi) + G_{5}'(\phi)/2$  \\ 
  \hline
  2.a.ii.1.a.i.1.b.i.1 & $g'(\phi)\neq0$  \\ 
   \hline
   2.a.ii.1.a.i.1.b.i.1.a & $c_2\neq-9/5$  \\ 
   \hline
    2.a.ii.1.a.i.1.b.i.1.b & $c_2=-9/5$  \\ 
   \hline
   2.a.ii.1.a.i.1.b.i.2 & $g'(\phi)=0$  \\ 
   \hline
   2.a.ii.1.a.i.1.b.ii & $G_{4b}(\phi)=3 g(\phi)^2/4 g'(\phi) +G_5'(\phi)/2$  \\ 
  \hline
   2.a.ii.1.a.i.1.b.ii.1 & $G_{{\rm Tele} 2}(\phi,T,I_2)=-G_{{\rm Tele} 1b}(\phi,I_2)+G_{{\rm Tele} 2a}(\phi,T)+3 I_2 T g(\phi)^3/(4g'(\phi)^2) +2 I_2 G_{4a}'(\phi)+$
   
   \vspace{1.5mm}
   + $ I_2 g(\phi) (-4X g'(\phi)-4h'(\phi)+T G_5'(\phi)+2 G_{2,X}(\phi,X)+2 G_{{\rm Tele} 1a,X}(\phi,X))\,/\,2\,g'(\phi)$  \\ 
  \hline
  2.a.ii.1.a.i.1.b.ii.1.a & $ h(\phi)\neq c_2 + c_3 g(\phi)$  \\ 
  \hline
  2.a.ii.1.a.i.1.b.ii.1.b & $h(\phi)= c_2 + c_3 g(\phi)$  \\ 
  \hline
  2.a.ii.1.a.i.1.b.ii.2 & $G_{{\rm Tele} 2}(\phi,T,I_2)=-G_{{\rm Tele} 1b}(\phi,I_2)+G_{{\rm Tele} 2a}(\phi,T)+3 I_2 T g(\phi)^3/(4g'(\phi)^2) +2 I_2 G_{4a}'(\phi)+$
   
   \vspace{1.5mm}
   + $ I_2 g(\phi) (-4X g'(\phi)-4h'(\phi)+T G_5'(\phi)+2 G_{2,X}(\phi,X)+2 G_{{\rm Tele} 1a,X}(\phi,X))\,/\,2\,g'(\phi)$  \\ 
  \hline
  2.a.ii.1.a.i.2 & $G_5'(\phi)=0$ \\
  \hline
  2.a.ii.1.a.i.2.a & $G_{{\rm Tele} 1}(\phi,X,I_2)\neq G_{{\rm Tele 1a}}(\phi,X) + G_{{\rm Tele}1b}(\phi,I_2)+2 I_2 X G_{4b}'(\phi)$  \\ 
  \hline
   2.a.ii.1.a.i.2.a.i & $G_{{\rm Tele} 1}\neq -I_2 X g(\phi)+G_{{\rm Tele} 1a}(\phi,X)+G_{{\rm Tele} 1b}(\phi,I_2)+4 I_2 X G_{4b}(\phi) g'(\phi)/3 g(\phi)+2 X I_2 G_{4b}'(\phi)$  \\ 
  \hline
   2.a.ii.1.a.i.2.a.ii & $G_{{\rm Tele} 1}= -I_2 X g(\phi)+G_{{\rm Tele} 1a}(\phi,X)+G_{{\rm Tele} 1b}(\phi,I_2)+4 I_2 X G_{4b}(\phi) g'(\phi)/3 g(\phi)+2 X I_2 G_{4b}'(\phi)$  \\ 
  \hline
   2.a.ii.1.a.i.2.b & $G_{{\rm Tele} 1}(\phi,X,I_2)= G_{{\rm Tele 1a}}(\phi,X) + G_{{\rm Tele}1b}(\phi,I_2)+2 I_2 X G_{4b}'(\phi)$  \\ 
  \hline
  2.a.ii.1.a.i.2.b.i & $g'(\phi)\neq0$  \\ 
  \hline
   2.a.ii.1.a.i.2.b.i.1 & $G_{{\rm Tele}2}(\phi,T,I_2) = 3 T X g(\phi)^2-3 T g(\phi)G_{4a}'(\phi)+$
   
   \vspace{1.5mm}
   + $2g'(\phi)(2TG_{4a}(\phi)-G_{{\rm Tele}1b}(\phi,I_2)+G_{{\rm Tele}2a}(\phi,I_2+3Tg(\phi)/(4g'(\phi))))$ \\ 
  \hline
   2.a.ii.1.a.i.2.b.i.1.a & $G_{{\rm Tele}2a}(\phi,T,I_2) = G_{{\rm Tele}2a1}(\phi)+2(I_2+3Tg(\phi)/(4g'(\phi)))/(9T)(9TG_{4a}'(\phi)+$
   
   \vspace{1.5mm}
   + $(6I_2^2-6I_2(I_2+3Tg(\phi)/(4g'(\phi)))+2(I_2+3Tg(\phi)/(4g'(\phi)))^2-9TX)g(\phi)-$
   
   \vspace{1.5mm}
   $-6h'(\phi)(-I_2+3Tg(\phi)/(4g'(\phi))) +3(-I_2+3Tg(\phi)/(4g'(\phi)))(G_{2,X}(\phi,X)+G_{{\rm Tele}1a,X}(\phi,X)) )$ \\ 
  \hline
   2.a.ii.1.a.i.2.b.i.1.a.i & $h(\phi) = c_2 + c_3 g(\phi)$ \\ 
  \hline
  2.a.ii.1.a.i.2.b.i.1.a.ii & $h(\phi) \neq c_2 + c_3 g(\phi)$ \\ 
  \hline
   2.a.ii.1.a.i.2.b.i.1.b &  $G_{{\rm Tele}2a}(\phi,T,I_2) \neq G_{{\rm Tele}2a1}(\phi)+2(I_2+3Tg(\phi)/(4g'(\phi)))/(9T)(9TG_{4a}'(\phi)+$
   
   \vspace{1.5mm}
   + $(6I_2^2-6I_2(I_2+3Tg(\phi)/(4g'(\phi)))+2(I_2+3Tg(\phi)/(4g'(\phi)))^2-9TX)g(\phi)-$
   
   \vspace{1.5mm}
   $-6h'(\phi)(-I_2+3Tg(\phi)/(4g'(\phi))) +3(-I_2+3Tg(\phi)/(4g'(\phi)))(G_{2,X}(\phi,X)+G_{{\rm Tele}1a,X}(\phi,X)) )$ \\ 
  \hline
  2.a.ii.1.a.i.2.b.i.2. & $G_{{\rm Tele}2}(\phi,T,I_2) \neq 3 T X g(\phi)^2-3 T g(\phi)G_{4a}'(\phi)+$
   
   \vspace{1.5mm}
   + $2g'(\phi)(2TG_{4a}(\phi)-G_{{\rm Tele}1b}(\phi,I_2)+G_{{\rm Tele}2a}(\phi,I_2+3Tg(\phi)/(4g'(\phi))))$ \\ 
  \hline
     2.a.ii.1.a.i.2.b.ii & $g'(\phi)=0$  \\ 
  \hline
   2.a.ii.1.a.i.2.b.ii.1 & $G_{4b}(\phi)=c_2/h'(\phi)$  \\ 
  \hline

  \end{tabular}
  \newpage

\begin{tabular}{ | m{4cm} | m{13cm}|| } 
  \hline
    2.a.ii.1.a.i.2.b.ii.2 & $G_{4b}(\phi)\neq c_2/h'(\phi)$  \\ 
  \hline
    2.a.ii.1.a.ii & $G_{{\rm Tele}2,{TI_2}}(\phi,T,I_2)= 0$  \\ 
  \hline
   2.a.ii.1.a.ii.1 & $G_5'(\phi)\neq0$  \\ 
  \hline
  2.a.ii.1.a.ii.1.a & $G_{{\rm Tele} 1},_{XI_2}\neq 2 G_{4b}'$  \\ 
  \hline
  2.a.ii.1.a.ii.1.a.i & $3 g^2(\phi)+2 g'(\phi) ( -2 G_{4b}(\phi)+G_{5}'(\phi) )+ g(\phi)( -6 G_{4b}'(\phi)+ 3 G_{{\rm Tele} 1}^{(0,1,1)}(\phi,X,I_2))\neq0$  \\ 
  \hline
  2.a.ii.1.a.ii.1.a.i.1 & $G_{{\rm Tele} 2a},_{TT}\neq0$  \\ 
  \hline
  2.a.ii.1.a.ii.1.a.i.2 & $G_{{\rm Tele} 2a},_{TT}=0$  \\ 
  \hline
   2.a.ii.1.a.ii.1.a.i.2.a & $\eta_{\phi}(\phi)=0
  $  \\ 
  \hline
  2.a.ii.1.a.ii.1.a.i.2.a.i & $c_4\neq-1/2
  $  \\ 
  \hline
  2.a.ii.1.a.ii.1.a.i.2.a.i.1 & $G_{4b}(\phi)=0$  \\ 
  \hline
   2.a.ii.1.a.ii.1.a.i.2.a.i.2 & $G_{4b}(\phi)\neq0
  $  \\ 
  \hline
   2.a.ii.1.a.ii.1.a.i.2.a.i.2.a & $\eta_{I_2}(t,a,\phi,N,T,I_2)\neq 0$ \\
   \hline
   2.a.ii.1.a.ii.1.a.i.2.a.i.2.b & $\eta_{I_2}(t,a,\phi,N,T,I_2) = 0$  \\ 
  \hline
  2.a.ii.1.a.ii.1.a.i.2.a.ii & $c_4 = -1/2$  \\ 
  \hline
  2.a.ii.1.a.ii.1.a.i.2.a.ii.1 & $\eta_{a3c}\neq 0$  \\ 
  \hline
  2.a.ii.1.a.ii.1.a.i.2.a.ii.2 & $\eta_{a3c} = 0 $  \\ 
  \hline
    2.a.ii.1.a.ii.1.a.i.2.a.ii.2.a & $\eta_{I_2}(t,a,\phi,N,T,I_2)\neq 0$ \\
  \hline
  2.a.ii.1.a.ii.1.a.i.2.a.ii.2.b & $\eta_{I_2}(t,a,\phi,N,T,I_2) = 0$  \\ 
  \hline
  2.a.ii.1.a.ii.1.a.i.2.b & $\eta_{\phi}(\phi) \neq 0
  $  \\ 
  \hline
  2.a.ii.1.a.ii.1.a.i.2.b.i & $G_{{\rm Tele} 1c}\neq -1/2$  \\ 
  \hline
  2.a.ii.1.a.ii.1.a.i.2.b.i.1 & $G_{{\rm Tele} 1c}\neq -2$  \\ 
  \hline
  2.a.ii.1.a.ii.1.a.i.2.b.i.2 & $G_{{\rm Tele} 1c}= -2$  \\ 
  \hline
  2.a.ii.1.a.ii.1.a.i.2.b.ii & $G_{{\rm Tele} 1c}= -1/2$  \\ 
  \hline
  2.a.ii.1.a.ii.1.a.ii & $3 g^2(\phi)+2 g'(\phi) ( -2 G_{4b}(\phi)+G_{5}'(\phi) )+ g(\phi)( -6 G_{4b}'(\phi)+ 3 G_{{\rm Tele} 1,XI_2}(\phi,X,I_2))=0$  \\ 
  \hline
  2.a.ii.1.a.ii.1.a.ii.1 & $\eta_{\phi}(t,\phi,N,I_2)=\eta_{\phi}(t,\phi)$  \\ 
  \hline
  2.a.ii.1.a.ii.1.a.ii.1.a & $\eta_{\phi}(t,\phi)\neq 0$  \\ 
  \hline
  2.a.ii.1.a.ii.1.a.ii.1.b & $\eta_{\phi}(t,\phi)=0  $  \\ 
  \hline
  2.a.ii.1.a.ii.1.a.ii.1.b.i & $G_{{\rm Tele} 2a,{TT}}(\phi,T)\neq 0  $  \\ 
  \hline
  2.a.ii.1.a.ii.1.a.ii.1.b.i.1 & $G_{{\rm Tele} 2b}(\phi,I_2) = -G_{{\rm Tele} 1b}(\phi,I_2)+G_{{\rm Tele} 2b1}(\phi)+I_2 G_{{\rm Tele} 2b2}(\phi)$  \\ 
  \hline
  2.a.ii.1.a.ii.1.a.ii.1.b.i.2 & $G_{{\rm Tele} 2b}(\phi,I_2) \neq -G_{{\rm Tele} 1b}(\phi,I_2)+G_{{\rm Tele} 2b1}(\phi)+I_2 G_{{\rm Tele} 2b2}(\phi)$  \\
  \hline
   \end{tabular}
  
  \newpage
  
\begin{tabular}{ | m{4cm} | m{13cm}|| } 
  \hline
 
  2.a.ii.1.a.ii.1.a.ii.1.b.ii & $G_{{\rm Tele} 2a,{TT}}(\phi,T)=0$  \\ 
  \hline
  2.a.ii.1.a.ii.1.a.ii.1.b.ii.1 & $\eta_{a3c}=0 $  \\ 
  \hline
  2.a.ii.1.a.ii.1.a.ii.1.b.ii.2 & $\eta_{a3c} \neq 0 $  \\ 
  \hline
  2.a.ii.1.a.ii.1.a.ii.1.b.ii.2.a & $G_{4a}(\phi)\neq 0  $  \\ 
  \hline
  2.a.ii.1.a.ii.1.a.ii.1.b.ii.2.b & $G_{4a}(\phi) = 0  $  \\ 
  \hline
  2.a.ii.1.a.ii.1.a.ii.2 & $\eta_{\phi}(t,\phi,N,I_2)\neq \eta_{\phi}(t,\phi)$  \\ 
  \hline
  2.a.ii.1.a.ii.1.b & $G_{{\rm Tele} 1,{XI_2}}= 2 G_{4b}'(\phi)$  \\
  \hline
  2.a.ii.1.a.ii.1.b.i & $G_{4bc}\neq 1$ \\
  \hline
  2.a.ii.1.a.ii.1.b.i.1 & $G_{4bc}\neq -2$  \\ 
  \hline
  2.a.ii.1.a.ii.1.b.i.1.a & $G_{{\rm Tele} 2a,TT}(\phi,T)\neq 0$  \\ 
  \hline
  2.a.ii.1.a.ii.1.b.i.1.a.i & $G_{{\rm Tele} 1b,I_2I_2}(\phi,I_2)+G_{{\rm Tele} 2b,I_2I_2}(\phi,I_2)\neq0$  \\ 
  \hline
  2.a.ii.1.a.ii.1.b.i.1.a.ii & $G_{{\rm Tele} 1b,I_2I_2}(\phi,I_2)+G_{{\rm Tele} 2b,I_2I_2}(\phi,I_2) = 0$  \\ 
  \hline
  2.a.ii.1.a.ii.1.b.i.1.b & $G_{{\rm Tele} 2a,TT}(\phi,T)= 0$  \\ 
  \hline
  2.a.ii.1.a.ii.1.b.i.2 & $G_{4bc}= -2$  \\ 
  \hline
    2.a.ii.1.a.ii.1.b.i.2.a & $G_{{\rm Tele} 2a,TT}(\phi,T)\neq0$  \\ 
  \hline
  2.a.ii.1.a.ii.1.b.i.2.a.i & $G_{{\rm Tele} 1b,I_2I_2}(\phi,I_2)+G_{{\rm Tele} 2b,I_2I_2}(\phi,I_2)\neq0$  \\ 
  \hline
  
  2.a.ii.1.a.ii.1.b.i.2.a.ii & $G_{{\rm Tele} 1b,I_2I_2}(\phi,I_2)+G_{{\rm Tele} 2b,I_2I_2}(\phi,I_2) = 0$ \\ 
  \hline
  2.a.ii.1.a.ii.1.b.i.2.b & $G_{{\rm Tele} 2a,TT}(\phi,T)=0$  \\ 
  \hline
  2.a.ii.1.a.ii.1.b.i.2.b.i & $\eta_{\phi}(\phi) = 0$  \\ 
  \hline
  2.a.ii.1.a.ii.1.b.i.2.b.i.1 & $G_{{\rm Tele} 1b,I_2I_2}(\phi,I_2)+G_{{\rm Tele} 2b,I_2I_2}(\phi,I_2)\neq0$  \\ 
  \hline
  2.a.ii.1.a.ii.1.b.i.2.b.i.2 & $G_{{\rm Tele} 1b,I_2I_2}(\phi,I_2)+G_{{\rm Tele} 2b,I_2I_2}(\phi,I_2) = 0$  \\ 
  \hline
  2.a.ii.1.a.ii.1.b.i.2.b.ii & $\eta_{\phi}(\phi) \neq 0$  \\ 
  \hline
  2.a.ii.1.a.ii.1.b.i.2.b.ii.1 & $G_{{\rm Tele} 1b,I_2I_2}(\phi,I_2)+G_{{\rm Tele} 2b,I_2I_2}(\phi,I_2)\neq0$  \\ 
  \hline
  2.a.ii.1.a.ii.1.b.i.2.b.ii.2 & $G_{{\rm Tele} 1b,I_2I_2}(\phi,I_2)+G_{{\rm Tele} 2b,I_2I_2}(\phi,I_2)=0$  \\ 
  \hline
  2.a.ii.1.a.ii.1.b.i & $G_{4bc}= 1$  \\ 
  \hline
   2.a.ii.1.a.ii.2 & $G_5'(\phi)=0$  \\ 
  \hline
   2.a.ii.1.a.ii.2.a & $ G_{{\rm Tele} 1}(\phi,X,I_2) = -I_2 X g(\phi)+G_{{\rm Tele}1a}(\phi,X)+G_{{\rm Tele}1b}(\phi,I_2)+$
   
   \vspace{1.5mm}
   $+4 I_2 X G_{4b}(\phi)g'(\phi)/ 3 g(\phi) + 2 I_2 X G_{4b}'(\phi) $  \\ 
  \hline
  2.a.ii.1.a.ii.2.a.i & $G_{4bc}\neq0$  \\ 
  \hline
  2.a.ii.1.a.ii.2.a.ii & $ G_{4bc} = 0$  \\ 
  \hline
   2.a.ii.1.a.ii.2.b & $G_{{\rm Tele} 1}(\phi,X,I_2) \neq -I_2 X g(\phi)+G_{{\rm Tele}1a}(\phi,X)+G_{{\rm Tele}1b}(\phi,I_2)+$
   
   \vspace{1.5mm}
   $+4 I_2 X G_{4b}(\phi)g'(\phi)/ 3 g(\phi) + 2 I_2 X G_{4b}'(\phi)$  \\ 
  \hline
  \end{tabular}
  
   \newpage

\begin{tabular}{ | m{4cm} | m{13cm}|| } 
  \hline

   2.a.ii.1.b & $G_{4b}(\phi) = G_5'(\phi)/2$  \\ 
  \hline
   2.a.ii.1.b.i& $g(\phi)-G_5''(\phi)+G_{{\rm Tele} 1,X,I_2}(\phi,X,I_2)\neq 0$  \\ 
  \hline
  2.a.ii.1.b.i.1 & $G_{{\rm Tele} 2a,{TT}}\neq 0$  \\ 
  \hline
  2.a.ii.1.b.i.2 & $G_{{\rm Tele} 2a,{TT}} = 0$  \\ 
  \hline
  2.a.ii.1.b.i.2.a & $G_5'(\phi)=0$  \\ 
  \hline
  2.a.ii.1.b.i.2.b & $G_5'(\phi) \neq 0$  \\ 
  \hline
  2.a.ii.1.b.i.2.b.i & $G_{{\rm Tele} 2b,{I_2I_2}}(\phi,T)+G_{{\rm Tele} 1,{I_2I_2}}(\phi,X,I_2)\neq0$  \\ 
  \hline
  2.a.ii.1.b.i.2.b.ii & $G_{{\rm Tele} 2b,{I_2I_2}}(\phi,T)+G_{{\rm Tele} 1,{I_2I_2}}(\phi,X,I_2)= 0$  \\ 
  \hline
  2.a.ii.1.b.i.2.b.ii.1 & $G_{{\rm Tele} 2a1}(\phi) = 2 G_{4a}(\phi),\,\,\eta_{\phi } \neq 0$  \\ 
  \hline
  2.a.ii.1.b.i.2.b.ii.2 & $G_{{\rm Tele} 2a1}(\phi) = 2 G_{4a}(\phi),\,\,\eta_{\phi } = 0$  \\ 
  \hline
   2.a.ii.1.b.i.2.b.ii.3 & $G_{{\rm Tele} 2a1}(\phi) \neq 2 G_{4a}(\phi),\,\,\eta_{\phi } \neq 0$  \\ 
  \hline
  2.a.ii.1.b.ii & $g(\phi)-G_5''(\phi)+G_{{\rm Tele} 1,XI_2}(\phi,X,I_2) = 0$  \\ 
  \hline
  2.a.ii.1.b.ii.1 & $G_{{\rm Tele} 2a1}(\phi)\neq 2 G_{4a}(\phi)$  \\ 
  \hline
  2.a.ii.1.b.ii.2 & $G_{{\rm Tele} 2a1}(\phi)= 2 G_{4a}(\phi)$  \\ 
  \hline
  2.a.ii.1.b.ii.2.a & $G_{{\rm Tele} 1b,I_2I_2}(\phi,I_2)+G_{{\rm Tele} 2b,I_2I_2}(\phi,I_2) \neq 0$  \\ 
  \hline
  2.a.ii.1.b.ii.2.b & $G_{{\rm Tele} 1b,I_2I_2}(\phi,I_2)+G_{{\rm Tele} 2b,I_2I_2}(\phi,I_2) = 0$  \\ 
  \hline
  2.a.ii.2.& $g(\phi)=0$  \\ 
  \hline
  2.a.ii.2.a & $G_5'(\phi)\neq 0$  \\ 
  \hline
   2.a.ii.2.a.i & $G_{{\rm Tele} 1,{XI_2}}(\phi,X,I_2)\neq 2G_{4b}'(\phi)$  \\ 
  \hline
   2.a.ii.2.a.i.1 & $G_5'(\phi)\neq 2 G_{4b}(\phi)$  \\ 
  \hline
  2.a.ii.2.a.i.1.a & $G_{{\rm Tele} 2,{TI_2}}(\phi,T,I_2)\neq0$  \\ 
  \hline
  2.a.ii.2.a.i.1.a.i & $G_{{\rm Tele} 1,{XI_2I_2}} \neq 0$  \\ 
  \hline
  2.a.ii.2.a.i.1.a.ii & $G_{{\rm Tele} 1,{XI_2I_2}} = 0$  \\ 
  \hline
   2.a.ii.2.a.i.1.b & $G_{{\rm Tele} 2,{TI_2}}(\phi,T,I_2)=0$  \\ 
  \hline
   2.a.ii.2.a.i.2 & $G_5'(\phi) = 2 G_{4b}(\phi)$  \\ 
  \hline
   2.a.ii.2.a.i.2.a & $2G_{4a}(\phi)-G_{{\rm Tele} 2,{T}}(\phi,T,I_2)\neq0$  \\ 
  \hline
  2.a.ii.2.a.i.2.a.i & $G_{{\rm Tele} 2,{TI_2}}(\phi,T,I_2)\neq0$  \\ 
  \hline
  2.a.ii.2.a.i.2.a.i.1 & $h'(\phi)\neq0$  \\ 
  \hline
  2.a.ii.2.a.i.2.a.i.2 & $h'(\phi)=0$  \\ 
  \hline
  \end{tabular}
 
 \newpage

\begin{tabular}{ | m{3cm} | m{14cm}|| } 
  \hline 
  2.a.ii.2.a.i.2.a.ii & $G_{{\rm Tele} 2,{TI_2}}(\phi,T,I_2)=0$  \\ 
  \hline
  2.a.ii.2.a.i.2.b & $2G_{4a}(\phi)-G_{{\rm Tele} 2,{T}}(\phi,T,I_2)=0$  \\ 
  \hline
  2.a.ii.2.a.i.2.b.i & $G_{{\rm Tele} 1} = -G_{{\rm Tele} 4}+G_{{\rm Tele} 1a}+I_2(G'_{{\rm Tele} 3}+XG''_5(\phi))$  \\ 
  \hline
  2.a.ii.2.a.i.2.b.i.1 & $h'(\phi)\neq 0$  \\ 
  \hline
  2.a.ii.2.a.i.2.b.i.2 & $h'(\phi)= 0$  \\ 
  \hline
  2.a.ii.2.a.i.2.b.ii & $G_{{\rm Tele} 1}\neq -G_{{\rm Tele} 4}+G_{{\rm Tele} 1a}+I_2(G'_{{\rm Tele} 3}+XG''_5(\phi))$  \\ 
  \hline
  2.a.ii.2.a.i.2.b.ii.1 & $h'(\phi)\neq 0$  \\ 
  \hline
  2.a.ii.2.a.i.2.b.ii.2 & $h'(\phi)= 0$  \\ 
  \hline
   2.a.ii.2.a.ii & $G_{{\rm Tele} 1,{XI_2}}(\phi,X,I_2)= 2G'_{4b}(\phi)$  \\ 
  \hline
   2.a.ii.2.a.ii.1 & $G'_5(\phi)\neq 2 G_{4b}(\phi)$  \\ 
  \hline
  2.a.ii.2.a.ii.1.a & $h'(\phi)\neq 0$  \\ 
  \hline
   2.a.ii.2.a.ii.1.b & $h'(\phi)= 0$  \\ 
  \hline
  2.a.ii.2.a.ii.2 & $G'_5(\phi)=2 G_{4b}(\phi)$  \\ 
  \hline
  2.a.ii.2.a.ii.2.a & $G_{{\rm Tele} 1b,{I_2I_2}}(\phi,I_2)\neq - G_{{\rm Tele} 2,{I_2I_2}}(\phi,T,I_2)$  \\ 
  \hline
  2.a.ii.2.a.ii.2.b & $G_{{\rm Tele} 1b,{I_2I_2}}(\phi,I_2) = - G_{{\rm Tele} 2,{I_2I_2}}(\phi,T,I_2)$   \\ 
  \hline
  2.a.ii.2.a.ii.2.b.i & $G_{{\rm Tele} 2b,T}(\phi,T) \neq0$  \\ 
  \hline
  2.a.ii.2.a.ii.2.b.ii & $G_{{\rm Tele} 2b,T}(\phi,T) = 0$ \\ 
  \hline 
  2.a.ii.2.b & $G_5'(\phi)= 0 $  \\ 
  \hline
  2.a.ii.2.b.i & $G_{4b}(\phi)\neq0$  \\ 
  \hline
  2.a.ii.2.b.i.1 & $G_{{\rm Tele} 2,TT}(\phi,T,I_2) \neq 0$  \\ 
  \hline
  2.a.ii.2.b.i.2 & $G_{{\rm Tele} 2,TT}(\phi,T,I_2)=0$  \\ 
  \hline
  2.a.ii.2.b.ii & $G_{4b}(\phi)=0$  \\ 
  \hline
  2.a.ii.2.b.ii.1 & $G_{{\rm Tele} 2,T}(\phi,T,I_2)\neq 2 G_{4a}(\phi)$  \\ 
  \hline
  2.a.ii.2.b.ii.2 & $G_{{\rm Tele} 2,T}(\phi,T,I_2) =  2 G_{4a}(\phi)$  \\ 
  \hline\hline
  \end{tabular}
  
  \vspace{1.5cm}
  
  \begin{tabular}{ | m{4cm} | m{13cm}|| } 
  \hline
   
  \multicolumn{2}{|c|}{\mathversion{bold}$2$.\bf{b} :  \mathversion{bold}$ \mathbf{G_{{\rm Tele},{XI2}}(\phi,X,T,I_2) = 2 G_{4b}'(\phi)-g(\phi)}$}\\
 \hline
 2.b.i & $g(\phi)\neq 0$  \\ 
  \hline
  2.b.i.1 & $g'(\phi)\neq0$  \\ 
  \hline
  2.b.i.1.a & $G_{4b}(\phi)\neq G'_{5}(\phi)/2$  \\ 
  \hline
  2.b.i.1.a.i & $3 G'_5(\phi)-2G_{4b}(\phi)+2 G_{{\rm Tele} 1,{XT}}(\phi,X,T)\neq0$  \\ 
  \hline
  2.b.i.1.a.i.1 & $\eta_{4cc}\neq-2$  \\ 
  \hline
  2.b.i.1.a.i.2 & $\eta_{4cc}=-2$  \\ 
  \hline
  2.b.i.1.a.i.2.a & $h(\phi)=c_5+c_4 g(\phi)$  \\ 
  \hline
  2.b.i.1.a.i.2.b & $h(\phi)\neq c_5+c_4 g(\phi)$  \\ 
  \hline
  2.b.i.1.a.ii & $3 G'_5(\phi)-2G_{4b}(\phi)+2 G_{{\rm Tele} 1,{XT}}(\phi,X,T)=0$  \\ 
  \hline
  2.b.i.1.a.ii.1 & $G_{4b}(\phi)\neq0$  \\ 
  \hline
  2.b.i.1.a.ii.1.a & $\eta_{a4c}\neq0$  \\ 
  \hline
  2.b.i.1.a.ii.1.a.i & $\eta_{a4c}=-3/2$  \\ 
  \hline
  2.b.i.1.a.ii.1.a.ii & $\eta_{a4c}\neq -3/2$  \\ 
  \hline
  2.b.i.1.a.ii.1.b & $\eta_{a4c}=0$  \\ 
  \hline
  2.b.i.1.a.ii.1.b.i & $-2G_{4a}(\phi)+G_{{\rm Tele} 1b,T}(\phi,T)+G_{{\rm Tele}2,T}(\phi,T,I_2)\neq0$  \\ 
  \hline
  2.b.i.1.a.ii.1.b.ii & $-2G_{4a}(\phi)+G_{{\rm Tele} 1b,T}(\phi,T)+G_{{\rm Tele}2,T}(\phi,T,I_2)=0$  \\ 
  \hline
     2.b.i.1.a.ii.2 & $G_{4b}(\phi)=0$  \\ 
  \hline
  2.b.i.1.a.ii.2.a & $\eta_{a3c}\neq -2$  \\ 
  \hline
  2.b.i.1.a.ii.2.b & $\eta_{a3c}= -2$  \\ 
  \hline
  2.b.i.1.b & $G_{4b}(\phi)= G'_{5}(\phi)/2$  \\ 
  \hline
  2.b.i.1.b.i & $G'_{5}(\phi)\neq G_{{\rm Tele} 1,XT}(\phi,X,T)$  \\ 
  \hline
   2.b.i.1.b.i.1 & $\eta_{a4c}\neq0$  \\ 
  \hline
  2.b.i.1.b.i.2 & $\eta_{a4c}=0$  \\ 
  \hline
  2.b.i.1.b.ii & $G'_{5}(\phi) = G_{{\rm Tele} 1,XT}(\phi,X,T)$  \\ 
  \hline
  2.b.i.1.b.ii.1 & $2 G_{4a}(\phi)-G_{{\rm Tele} 1b,T}(\phi,T)-G_{{\rm Tele} 2,T}(\phi,T,I_2)\neq0$  \\ 
  \hline
  2.b.i.1.b.ii.2 & $2 G_{4a}(\phi)-G_{{\rm Tele} 1b,T}(\phi,T)-G_{{\rm Tele} 2,T}(\phi,T,I_2)=0$  \\ 
  \hline
  2.b.i.1.b.ii.2.a & $G_{{\rm Tele} 2a,{I_2I_2}}(\phi,I_2)\neq0
  $  \\ 
  \hline
  2.b.i.1.b.ii.2.b & $G_{{\rm Tele} 2a,{I_2I_2}}(\phi,I_2)=0
  $  \\ 
  \hline
  \end{tabular}
  
  \newpage
  
  \begin{tabular}{ | m{4cm} | m{13cm}|| } 
  \hline

  2.b.i.1.b.ii.2.b.i & $\eta_{a4c}\neq0
  $  \\ 
  \hline
  2.b.i.1.b.ii.2.b.i.1 & $h(\phi)\neq c_5+c_4 g(\phi)  $  \\ 
  \hline
  2.b.i.1.b.ii.2.b.i.2 & $h(\phi) = c_5+c_4 g(\phi) $  \\ 
  \hline
  2.b.i.1.b.ii.2.b.ii & $\eta_{a4c}=0
  $  \\ 
  \hline
   2.b.i.1.b.ii.2.b.ii.1 & $G_{{\rm Tele} 2a2}(\phi)=c_7 g(\phi)+2 G'_{4a}(\phi)$  \\ 
  \hline
  2.b.i.1.b.ii.2.b.ii.1.a & $h(\phi) \neq c_4+G_{{\rm Tele}2a2c}g(\phi)  $  \\ 
  \hline
  2.b.i.1.b.ii.2.b.ii.1.b & $h(\phi) = c_4+G_{{\rm Tele}2a2c}g(\phi)$  \\ 
  \hline
  2.b.i.1.b.ii.2.b.ii.2 & $G_{{\rm Tele} 2a2}(\phi) \neq c_7 g(\phi)+2 G'_{4a}(\phi)$  \\ 
  \hline
   2.b.i.2 & $g'(\phi)= 0$  \\ 
  \hline
  2.b.i.2.a & $3 G'_{5}(\phi)-2(G_{4b}(\phi)+G_{{\rm Tele} 1,XT}(\phi,X,T))\neq 0$  \\ 
  \hline
  2.b.i.2.a.i & $G'_{5}(\phi) \neq G_{{\rm Tele} 1,XT}(\phi,X,T)$  \\ 
  \hline
  2.b.i.2.a.ii & $G'_{5}(\phi) = G_{{\rm Tele} 1,XT}(\phi,X,T)$  \\ 
  \hline
  2.b.i.2.b & $3 G'_{5}(\phi)-2(G_{4b}(\phi)+G_{{\rm Tele} 1,XT}(\phi,X,T))=0$  \\ 
  \hline
    2.b.i.2.b.i & $G'_{5}(\phi)\neq 2 G_{4b}(\phi)$  \\ 
  \hline
  2.b.i.2.b.i.1 & $G_{{\rm Tele} 1b,TT}(\phi,T)+G_{{\rm Tele} 2,{TT}}(\phi,T,I_2)\neq 0$  \\ 
  \hline
  2.b.i.2.b.i.1.a & $G_{{\rm Tele} 2}(\phi,T,I_2) \neq G_{{\rm Tele} 2a}(\phi,T)+I_2 G_{{\rm Tele} 2b}(\phi,T)$  \\ 
  \hline
  2.b.i.2.b.i.1.b & $G_{{\rm Tele} 2}(\phi,T,I_2) = G_{{\rm Tele} 2a}(\phi,T)+I_2 G_{{\rm Tele} 2b}(\phi,T)$  \\ 
  \hline
  2.b.i.2.b.i.2 & $G_{{\rm Tele} 1b,TT}(\phi,T)+G_{{\rm Tele} 2,{TT}}(\phi,T,I_2) =0$  \\ 
  \hline
  2.b.i.2.b.i.2.a & $G_{4a}(\phi) \neq 0$  \\ 
  \hline
  2.b.i.2.b.i.2.b & $G_{4a}(\phi)=0 $  \\ 
  \hline
  2.b.i.2.b.ii & $G'_{5}(\phi)= 2 G_{4b}(\phi)$  \\ 
  \hline
   2.b.i.2.b.ii.1 & $h'(\phi)\neq 0$  \\ 
  \hline
   2.b.i.2.b.ii.1.a & $G_{{\rm Tele} 2,{TI_2}}(\phi,T,I_2)\neq 0$  \\ 
  \hline
  2.b.i.2.b.ii.1.b & $G_{{\rm Tele} 2,{TI_2}}(\phi,T,I_2)= 0$  \\ 
  \hline
  2.b.i.2.b.ii.1.b.i & $G_{{\rm Tele} 2b,{I_2I_2}}(\phi,I_2) \neq 0
  $  \\ 
  \hline
  2.b.i.2.b.ii.1.b.i.1 & $G_{{\rm Tele} 1b,{TT}}(\phi,T)+G_{{\rm Tele} 2a,{TT}}(\phi,T) \neq 0  $  \\ 
  \hline
  2.b.i.2.b.ii.1.b.i.2 & $G_{{\rm Tele} 1b,{TT}}(\phi,T)+G_{{\rm Tele} 2a,{TT}}(\phi,T)  = 0  $  \\ 
  \hline
  2.b.i.2.b.ii.1.b.i.2.a & $G_5'(\phi)\neq 0  $  \\ 
  \hline
   2.b.i.2.b.ii.1.b.i.2.b & $G_5'(\phi) =0   $  \\ 
  \hline
  \end{tabular}
  
   \newpage
  
  \begin{tabular}{ | m{4cm} | m{13cm}|| } 
  \hline
  2.b.i.2.b.ii.1.b.ii & $G_{{\rm Tele} 2b,{I_2I_2}}(\phi,I_2)=0
  $  \\ 
  \hline
   2.b.i.2.b.ii.1.b.ii.1 & $G_{{\rm Tele} 1b,{TT}}(\phi,T)+G_{{\rm Tele} 2a,{TT}}(\phi,T)\neq0
  $  \\ 
  \hline
   2.b.i.2.b.ii.1.b.ii.2 & $G_{{\rm Tele} 1b,{TT}}(\phi,T)+G_{{\rm Tele} 2a,{TT}}(\phi,T) = 0  $  \\ 
  \hline
   2.b.i.2.b.ii.1.b.ii.2.a & $G_5'(\phi) \neq 0  $  \\ 
  \hline
    2.b.i.2.b.ii.1.b.ii.2.b & $G_5'(\phi) = 0  $  \\ 
  \hline
   2.b.i.2.b.ii.2 & $h'(\phi) = 0$  \\ 
  \hline
  2.b.i.2.b.ii.2.a & $G_{{\rm Tele} 2,{I_2I_2}}(\phi,T,I_2) \neq0$  \\ 
  \hline
   2.b.i.2.b.ii.2.a.i & $G_5'(\phi)\neq 0$  \\ 
  \hline
   2.b.i.2.b.ii.2.a.ii & $G_5'(\phi) = 0$  \\ 
  \hline
  2.b.i.2.b.ii.2.b & $G_{{\rm Tele} 2,{I_2I_2}}(\phi,T,I_2)=0$  \\ 
  \hline
   2.b.i.2.b.ii.2.b.i & $G_{{\rm Tele} 2b,T}(\phi,T)\neq0$  \\ 
  \hline
   2.b.i.2.b.ii.2.b.ii & $G_{{\rm Tele} 2b,T}(\phi,T) = 0$  \\ 
  \hline
  2.b.i.2.b.ii.2.b.ii.1 & $G_{{\rm Tele} 1b,{TT}}(\phi,T)+G_{{\rm Tele} 2a,{TT}}(\phi,T)\neq0$
  \\ 
  \hline
  2.b.i.2.b.ii.2.b.ii.2 & $G_{{\rm Tele} 1b,{TT}}(\phi,T)+G_{{\rm Tele} 2a,{TT}}(\phi,T) = 0$  \\ 
  \hline\hline
     2.b.ii & $g(\phi)= 0$  \\ 
  \hline
   2.b.ii.1 & $G_{4b}(\phi)\neq G'_5(\phi)/2$  \\ 
  \hline
  2.b.ii.1.a & $G'_{4a}(\phi) \neq G_{{\rm Tele} 2,{I_2}}(\phi,X,I_2)/2 $ \\ 
  \hline
   2.b.ii.1.a.i & $3 G'_5(\phi) - 2 G_{4b}(\phi)-2 G_{{\rm Tele} 1,XT}(\phi,X,T)\neq0 $ \\ 
  \hline
   2.b.ii.1.a.ii & $3 G'_5(\phi) - 2 G_{4b}(\phi)-2 G_{{\rm Tele} 1,XT}(\phi,X,T)=0 $ \\ 
  \hline
   2.b.ii.1.a.ii.1 & $h'(\phi) \neq 0 $ \\ 
  \hline
   2.b.ii.1.a.ii.1.a & $G_{{\rm Tele} 2,{I_2I_2}}(\phi,T,I_2)\neq0 $ \\ 
  \hline
   2.b.ii.1.a.ii.1.b & $G_{{\rm Tele} 2,{I_2I_2}}(\phi,T,I_2) = 0 $ \\ 
  \hline
  2.b.ii.1.a.ii.1.b.i & $G_{{\rm Tele} 2b,{T}}(\phi,T)\neq0$
   \\ 
  \hline
   2.b.ii.1.a.ii.1.b.ii & $G_{{\rm Tele} 2b,{T}}(\phi,T)=0$
   \\ 
  \hline
      2.b.ii.1.a.ii.1.b.ii.1 & $G_{{\rm Tele} 1b,{TT}}(\phi,T)+G_{{\rm Tele} 2a,{TT}}(\phi,T) \neq 0$
   \\ 
  \hline
   2.b.ii.1.a.ii.1.b.ii.2 & $G_{{\rm Tele} 1b,{TT}}(\phi,T)+G_{{\rm Tele} 2a,{TT}}(\phi,T)=0$
   \\ 
  \hline
   2.b.ii.1.a.ii.2 & $h'(\phi) = 0 $ \\ 
  \hline
   2.b.ii.1.a.ii.2.a & $G_{{\rm Tele} 2,{I_2I_2}}(\phi,T,I_2)\neq0 $ \\ 
  \hline
   2.b.ii.1.a.ii.2.b & $G_{{\rm Tele} 2,{I_2I_2}}(\phi,T,I_2) = 0 $ \\ 
  \hline

  \end{tabular}
  
   \newpage
  
  \begin{tabular}{ | m{4cm} | m{13cm}|| } 
  \hline
   2.b.ii.1.a.ii.2.b.i & $G_{{\rm Tele} 2b,T}(\phi,T)\neq0$
    \\ 
  \hline
   2.b.ii.1.a.ii.2.b.ii & $G_{{\rm Tele} 2b,T}(\phi,T)=0,$   \\ 
  \hline
   2.b.ii.1.a.ii.2.b.ii.1 & $G_{{\rm Tele} 1b,{TT}}(\phi,T)+G_{{\rm Tele} 2a,{TT}}(\phi,T)\neq0$
    \\ 
  \hline
   2.b.ii.1.a.ii.2.b.ii.2 & $G_{{\rm Tele} 1b,{TT}}(\phi,T)+G_{{\rm Tele} 2a,{TT}}(\phi,T) = 0$   \\ 
  \hline
   2.b.ii.1.b & $G'_{4a}(\phi) = G_{{\rm Tele} 2,{I_2}}(\phi,X,I_2)/2$ \\ 
  \hline
   2.b.ii.1.b.i & $-G'_5 (\phi)+G_{{\rm Tele} 1,{XT}}(\phi,X,T)\neq0 $ \\ 
  \hline
  2.b.ii.1.b.i.1 & $3 G'_5(\phi) -2 G_{4b}(\phi) - 2 G_{{\rm Tele} 1,{XT}}(\phi,X,T)\neq0 $ \\ 
  \hline
   2.b.ii.1.b.i.2 & $3 G'_5(\phi) -2 G_{4b}(\phi) - 2 G_{{\rm Tele} 1,{XT}}(\phi,X,T)= 0 $ \\ 
  \hline
   2.b.ii.1.b.i.2.a & $G_{{\rm Tele} 1b,{TT}}(\phi,T)+G_{{\rm Tele} 2a,{TT}}(\phi,T)\neq0$ \\ 
  \hline
   2.b.ii.1.b.i.2.a.i & $h'(\phi) \neq 0$ \\ 
  \hline
   2.b.ii.1.b.i.2.a.ii & $h'(\phi) = 0$ \\ 
  \hline
    2.b.ii.1.b.i.2.b & $G_{{\rm Tele} 1b,{TT}}(\phi,T)+G_{{\rm Tele} 2a,{TT}}(\phi,T)=0$
    \\ 
  \hline
   2.b.ii.1.b.i.2.b.i & $h'(\phi) \neq 0$    \\ 
  \hline
   2.b.ii.1.b.i.2.b.ii & $h'(\phi) = 0$    \\ 
  \hline
   2.b.ii.1.b.ii & $-G'_5 (\phi)+G_{{\rm Tele} 1,{XT}}(\phi,X,T) = 0 $ \\ 
  \hline
   2.b.ii.1.b.ii.1 & $G_{4b}(\phi)\neq0 $ \\ 
  \hline
   2.b.ii.1.b.ii.1.a & $h'(\phi) \neq 0 $ \\ 
  \hline
   2.b.ii.1.b.ii.1.b & $h'(\phi) = 0$ \\ 
  \hline
   2.b.ii.1.b.ii.2 & $G_{4b}(\phi)=0 $ \\ 
  \hline
   2.b.ii.1.b.ii.2.a & $-2h'(\phi)+G_{2,X}(\phi,X)+G_{{\rm Tele} 1a,X}(\phi,X) \neq 0 $ \\ 
  \hline
   2.b.ii.1.b.ii.2.a.i & $G_{{\rm Tele} 1b,{TT}}(\phi,T)+G_{{\rm Tele} 2a,{TT}}(\phi,T)\neq0$
   \\ 
  \hline
   2.b.ii.1.b.ii.2.a.ii & $G_{{\rm Tele} 1b,{TT}}(\phi,T)+G_{{\rm Tele} 2a,{TT}}(\phi,T)=0 $   \\ 
  \hline
   2.b.ii.1.b.ii.2.b & $-2h'(\phi)+G_{2,X}(\phi,X)+G_{{\rm Tele} 1a,X}(\phi,X)=0 $ \\ 
  \hline
   2.b.ii.2 & $G_{4b}(\phi)= G'_5(\phi)/2$  \\ 
  \hline
   2.b.ii.2.a & $-G'_5 (\phi)+G_{{\rm Tele} 1,{XT}}(\phi,X,T) \neq 0$  \\ 
  \hline
   2.b.ii.2.a.i & $2 G'_{4a}(\phi)\neq G_{{\rm Tele} 2,{I_2}}(\phi,T,I_2)$  \\ 
  \hline
    2.b.ii.2.a.i.1 & $\eta_{a1,{\phi}}(a,\phi)\neq0$  \\ 
  \hline
   2.b.ii.2.a.i.2 & $\eta_{a1,{\phi}}(a,\phi)=0$  \\ 
  \hline
   2.b.ii.2.a.i.2.a & $h'(\phi)\neq 0$  \\ 
  \hline
   2.b.ii.2.a.i.2.a.i & $G_{{\rm Tele} 2,{I_2I_2}} (\phi,T,I_2) \neq0$  \\ 
  \hline
    2.b.ii.2.a.i.2.a.ii & $G_{{\rm Tele} 2,{I_2I_2}} (\phi,T,I_2)=0$ \\
    \hline
  \end{tabular}
  
   \newpage
  
  \begin{tabular}{ | m{4cm} | m{13cm}|| } 
  \hline
    2.b.ii.2.a.i.2.a.ii.1 & $G_{{\rm Tele} 2b,T}(\phi,T)\neq0$     \\
    \hline
     2.b.ii.2.a.i.2.a.ii.2 & $G_{{\rm Tele} 2b,T}(\phi,T)=0$     \\
    \hline
     2.b.ii.2.a.i.2.a.ii.2.a & $G_{{\rm Tele} 2a,TT}(\phi,T)+ G_{{\rm Tele}1,TT}(\phi,X,T)\neq0$     \\
    \hline
     2.b.ii.2.a.i.2.a.ii.2.b & $G_{{\rm Tele} 2a,TT}(\phi,T)+ G_{{\rm Tele}1,TT}(\phi,X,T) = 0 $     \\
    \hline
     2.b.ii.2.a.i.2.b & $h'(\phi) = 0$  \\ 
  \hline
   2.b.ii.2.a.i.2.b.i & $G_{{\rm Tele}2,I_2I_2}(\phi,T,I_2)\neq0$  \\ 
  \hline
   2.b.ii.2.a.i.2.b.ii & $G_{{\rm Tele}2,I_2I_2}(\phi,T,I_2) = 0$  \\ 
  \hline
   2.b.ii.2.a.i.2.b.ii.1 & $G_{{\rm Tele} 2b,T}(\phi,T)\neq 0$     \\ 
  \hline
   2.b.ii.2.a.i.2.b.ii.2 & $G_{{\rm Tele} 2b,T}(\phi,T)=0 $      \\ 
  \hline
   2.b.ii.2.a.i.2.b.ii.2.a & $G_{{\rm Tele} 2b1}(\phi) \neq 2 G'_{4a}(\phi)$
     \\ 
  \hline
   2.b.ii.2.a.i.2.b.ii.2.b & $G_{{\rm Tele} 2b1}(\phi) = 2 G'_{4a}(\phi)$
     \\ 
  \hline
    2.b.ii.2.a.ii & $2 G'_{4a}(\phi)= G_{{\rm Tele} 2,{I_2}}(\phi,T,I_2)$  \\ 
  \hline
    2.b.ii.2.a.ii.1 & $h'(\phi)\neq 0$  \\ 
  \hline
   2.b.ii.2.a.ii.2 & $h'(\phi) = 0$  \\ 
  \hline
    2.b.ii.2.a.ii.2.a & $G_{{\rm Tele}1}(\phi,X,T) = 2 T G_{4a}(\phi) -G_{{\rm Tele} 2a}(\phi,T) + G_{{\rm Tele} 1a}(\phi,X) + T X  G'_5(\phi)$  \\ 
  \hline
    2.b.ii.2.a.ii.2.b & $G_{{\rm Tele}1}(\phi,X,T) \neq 2 T G_{4a}(\phi) -G_{{\rm Tele} 2a}(\phi,T) + G_{{\rm Tele} 1a}(\phi,X) + T X  G'_5(\phi)$  \\ 
  \hline
   2.b.ii.2.b & $-G'_5 (\phi)+G_{{\rm Tele} 1,{XT}}(\phi,X,T) = 0$  \\ 
  \hline
   2.b.ii.2.b.i & $G_{{\rm Tele} 2}(\phi,T,I_2) = 2T G_{4a}(\phi) - G_{{\rm Tele} 1b}(\phi,T) + G_{{\rm Tele} 2a}(\phi,I_2)$  \\ 
  \hline
    2.b.ii.2.b.i.1 & $G_{{\rm Tele} 2a}(\phi,I_2) = G_{{\rm Tele} 2a1}(\phi) + 2 I_2 G'_{4a}(\phi)$  \\ 
  \hline
   2.b.ii.2.b.i.1.a & $4 h'(\phi) - T G'_5(\phi)-2 ( G_{2,X}(\phi,X) + G_{{\rm Tele} 1a,X}(\phi,X)) =  0$\\
  \hline
   2.b.ii.2.b.i.1.a.i & $h'(\phi) = 0$\\
  \hline
   2.b.ii.2.b.i.1.a.i.1 & $G_{2b}(\phi)+G_{{\rm Tele} 2a1}(\phi) + T G_{4a}(\phi)=0$\\
  \hline
  2.b.ii.2.b.i.1.a.i.2 & $G_{2b}(\phi)+G_{{\rm Tele} 2a1}(\phi) + T G_{4a}(\phi)\neq0$\\
  \hline
   2.b.ii.2.b.i.1.a.ii & $h'(\phi) \neq 0$\\
  \hline
  2.b.ii.2.b.i.1.b & $4 h'(\phi) - T G'_5(\phi)-2 ( G_{2,X}(\phi,X) + G_{{\rm Tele} 1a,X}(\phi,X)) \neq 0$\\
  \hline
   2.b.ii.2.b.i.2 & $G_{{\rm Tele} 2a}(\phi,I_2) \neq G_{{\rm Tele} 2a1}(\phi) + 2 I_2 G'_{4a}(\phi)$  \\ 
  \hline
   2.b.ii.2.b.ii & $G_{{\rm Tele} 2}(\phi,T,I_2) \neq 2T G_{4a}(\phi) - G_{{\rm Tele} 1b}(\phi,T) + G_{{\rm Tele} 2a}(\phi,I_2)$  \\ 
  \hline\hline
  \end{tabular}

\bibliographystyle{utphys}
\bibliography{references}

\end{document}